\documentclass[letterpaper,twocolumn,11pt,accepted=2021-11-06]{quantumarticle}

\pdfoutput=1
\usepackage[scaled]{beramono}
\usepackage[T1]{fontenc}
\usepackage[utf8]{inputenc}
\usepackage[english]{babel}
\usepackage{amsmath}
\usepackage{hyperref}
\usepackage{amssymb, amsmath}
\usepackage[numbers,sort&compress]{natbib}
\usepackage{cleveref}

\usepackage{colortbl}
\usepackage[table]{xcolor}
\usepackage{nonfloat}
\usepackage{enumitem}
\usepackage{enumitem}
\setitemize{noitemsep,topsep=2pt,parsep=0pt,partopsep=0pt}
\setenumerate{noitemsep,topsep=2pt,parsep=0pt,partopsep=0pt}

\usepackage{color}
\definecolor{deepblue}{HTML}{002AA0}
\definecolor{deepred}{rgb}{0.6,0,0}
\definecolor{deepgreen}{rgb}{0,0.5,0}
\definecolor{lightblue}{HTML}{F0F8FF}

\usepackage{listings}
\definecolor{codegreen}{rgb}{0,0.6,0}
\definecolor{codegray}{rgb}{0.5,0.5,0.5}
\definecolor{backcolor}{HTML}{ECEFF1}

\usepackage{xspace}

\usepackage{graphicx}
\graphicspath{{./graphics/}}

\lstdefinestyle{codeblock}{
	backgroundcolor=\color{backcolor},   
	commentstyle=\color{codegreen},
	keywordstyle=\color{deepblue},
	numberstyle=\tiny\color{codegray},
	stringstyle=\color{deepgreen},
	basicstyle=\footnotesize,
	escapechar=\¢,escapebegin=\color{purple}, % highlighting for decorator
	otherkeywords={with},
	breakatwhitespace=false,         
	breaklines=true,      
	lineskip=1pt,
	captionpos=b,                    
	keepspaces=true,
	language=Python,
	numbers=left,                    
	numbersep=6pt,                  
	showspaces=false,                
	showstringspaces=false,
	showtabs=false,                  
	tabsize=2,
	frame=single,
	framerule=0pt,
	basicstyle=\fontencoding{T1}\ttfamily\footnotesize
}
\lstdefinestyle{output}{
	backgroundcolor=\color{white},   
	commentstyle=\color{codegreen},
	keywordstyle=\color{deepblue},
	numberstyle=\tiny\color{white},
	stringstyle=\color{deepgreen},
	basicstyle=\footnotesize,
	escapechar=\¢,escapebegin=\color{purple}, % highlighting for decorator
	otherkeywords={with},
	breakatwhitespace=false,         
	breaklines=true,      
	lineskip=1pt,
	captionpos=b,                    
	keepspaces=true,
	language=Python,
	numbers=none,                    
	numbersep=6pt,                  
	showspaces=false,                
	showstringspaces=false,
	showtabs=false,                  
	tabsize=2,
	frame=single,
	framerule=0pt,
	basicstyle=\fontencoding{T1}\ttfamily\footnotesize
}
\newcommand{\pyl}[1]{\lstinline!#1!}

\makeatletter
\newcommand*{\lstitem}[1][]{%
  \setbox0\hbox\bgroup
    \patchcmd{\lst@InlineM}{\@empty}{\@empty\egroup\item[\usebox0]\leavevmode\ignorespaces}{}{}%
    \lstinline[#1]%
}
\makeatother

\newcommand{\Aop}{{\hat{A} }}
\newcommand{\Bop}{{\hat{B} }}
\newcommand{\Hop}{{\hat{H} }}
\newcommand{\Pop}{{\hat{P} }}
\newcommand{\Vop}{{\hat{V} }}
\newcommand{\Xop}{{\hat{X} }}
\newcommand{\aop}{{\hat{a} }}
\newcommand{\dg}{^{\dagger}}
\newcommand{\noper}{{ \hat{n} }}
\newcommand{\nop}{{\hat{n} }}
\newcommand{\phiop}{{\hat{\phi} }}
\newcommand{\scqubits}{\textsf{scqubits}\xspace}
\newcommand{\thetaop}{{\hat{\theta} }}
\newcommand{\varphiop}{{\hat{\varphi} }}
\newcommand{\zetaop}{{\hat{\zeta} }}

\begin{document}
\title{scqubits: a Python package for superconducting qubits}

\author{Peter Groszkowski}
\affiliation{Pritzker School for Molecular Engineering, University of Chicago, 5640 South Ellis Avenue, Chicago, IL 60637, USA}

\author{Jens Koch}
\email{jens-koch@northwestern.edu}
\orcid{0000-0002-5047-631X}
\affiliation{Department of Physics and Astronomy, Northwestern University, Evanston, IL 60208, USA}

\begin{abstract}
\scqubits is an open-source Python package for simulating and analyzing superconducting circuits. It provides convenient routines to obtain energy spectra of common superconducting qubits, such as the transmon, fluxonium, flux, cos(2$\phi$) and the 0-$\pi$ qubit. \scqubits also features a number of options for visualizing the computed spectral data, including plots of
energy levels as a function of external parameters, display of matrix elements of various operators as well as means to easily plot qubit wavefunctions. Many of these tools are not limited to single qubits, but extend to composite Hilbert spaces consisting of coupled superconducting qubits and
harmonic (or weakly anharmonic) modes. The library provides an extensive suite of methods for estimating qubit coherence times due to a variety of commonly considered noise channels.  While all functionality of \scqubits can be accessed programatically, the package also implements GUI-like widgets that, with a few clicks can help users both create relevant Python objects, as well as explore their properties through various plots. When applicable, the library harnesses the computing power of multiple cores via multiprocessing. \scqubits further exposes a direct interface to the Quantum Toolbox in Python (QuTiP) package, allowing the user to efficiently leverage QuTiP's proven capabilities for simulating time evolution.

%arxiv submission version
%$\textbf{scqubits}$ is an open-source Python package for simulating and analyzing superconducting circuits. It provides convenient routines to obtain energy spectra of common superconducting qubits, such as the transmon, fluxonium, flux, cos(2$\phi$) and the 0-$\pi$ qubit. $\textbf{scqubits}$ also features a number of options for visualizing the computed spectral data, including plots of energy levels as a function of external parameters, display of matrix elements of various operators as well as means to easily plot qubit wavefunctions. Many of these tools are not limited to single qubits, but extend to composite Hilbert spaces consisting of coupled superconducting qubits and harmonic (or weakly anharmonic) modes. The library provides an extensive suite of methods for estimating qubit coherence times due to a variety of commonly considered noise channels. While all functionality of $\textbf{scqubits}$ can be accessed programatically, the package also implements GUI-like widgets that, with a few clicks can help users both create relevant Python objects, as well as explore their properties through various plots. When applicable, the library harnesses the computing power of multiple cores via multiprocessing. $\textbf{scqubits}$ further exposes a direct interface to the Quantum Toolbox in Python (QuTiP) package, allowing the user to efficiently leverage QuTiP's proven capabilities for simulating time evolution.

\end{abstract}

\renewcommand{\ttdefault}{fvm}
%% main text
\section{Introduction}
\label{sec:intro}

Superconducting qubits \cite{Devoret2004a,You2005,Clarke2008,Krantz2019} have secured the rank of one of the most promising and widely researched hardware architectures for quantum information processing. All devices in this category are relatively simple circuits, and display genuine quantum properties such as discrete energy spectra and quantum-coherent time evolution. Coupling superconducting qubits to external electromagnetic fields enables the realization of gate operations as well as quantum-state readout using the framework of circuit quantum electrodynamics (circuit QED) \cite{Blais2004,Wallraff2004,Blais2021}.

The computation of energy spectra, eigenstates, and matrix elements of relevant operators is a key prerequisite for the design and fabrication of superconducting qubits, as well as for the quantitative analysis of experimental data collected in state-of-the-art experiments. Circuit quantization \cite{Devoret1995,Burkard2004,Vool2017} provides the systematic framework for deriving the Hamiltonian operator which describes a given circuit mathematically. However, with the exception of the simple LC oscillator and the limiting behavior of nonlinear circuits in specific parameter regimes, computation of qubit spectra cannot be accomplished analytically, but rather requires numerical solution of Hermitian eigenvalue problems.

The \scqubits package provides a user-friendly, object-oriented Python library of the most common superconducting qubits. It facilitates automatic construction of circuit Hamiltonians in an appropriate basis, provides high-level routines for finding eigenenergies, eigenstates, and matrix elements, and allows the user to quickly visualize these quantities as a function of external parameters. While the \scqubits routines for numerics and plotting rely heavily on NumPy, SciPy, and Matplotlib, no detailed knowledge of these internals is required from the user to employ \scqubits efficiently. In this way, the package helps lower initial barriers encountered by new researchers entering the field, and can be easily integrated for educational purposes. At the same time, the library aims to fulfill the role of a unifying tool for expert workers in the field, enabling quick and efficient investigations of a wide variety of circuit-QED systems, and the interactive exploration of their behavior in different parameter regimes. 

The analysis of interesting circuit-QED systems invariably involves the consideration of coupled systems, composed of qubits on one hand, and harmonic modes on the other hand (realized, for example, as on-chip transmission-line resonators or 3d cavities). The \scqubits library simplifies the setup of such composite Hilbert spaces and grants a seamless interface to the well-established QuTiP \cite{Johansson2012,Johansson2013a} framework which can be leveraged for the simulation of time evolution.

This paper is organized as follows. In the next section, we provide an overview of the library and its core functionality. In Sec.~\ref{sec:compositeHilberSpaces} we discuss how to construct and analyze composite Hilbert space systems that may consist of multiple qubits and/or resonators. Next, in Sec.~\ref{sec:sweeps} we present how \scqubits  lets users perform parameter sweeps and easily explore how properties of a composite system vary as circuit parameters or control fields change. In Sec.~\ref{sec:coherence} we review how \scqubits can be used to estimate coherence times of different qubits, and how those coherence times can be visualized. 
After that, in Sec.~\ref{sec:interactiveExploration} we provide a brief overview of the interactive exploration capabilities of the library, that allow users to study properties of various systems in ways that require very little actual programming. 
Finally, we summarize and conclude in Sec.~\ref{sec:conclusions}.

\section{Overview of the \scqubits library}
\label{sec:overview}
\begin{figure*}
	\centering
	\includegraphics[width=0.7\textwidth]{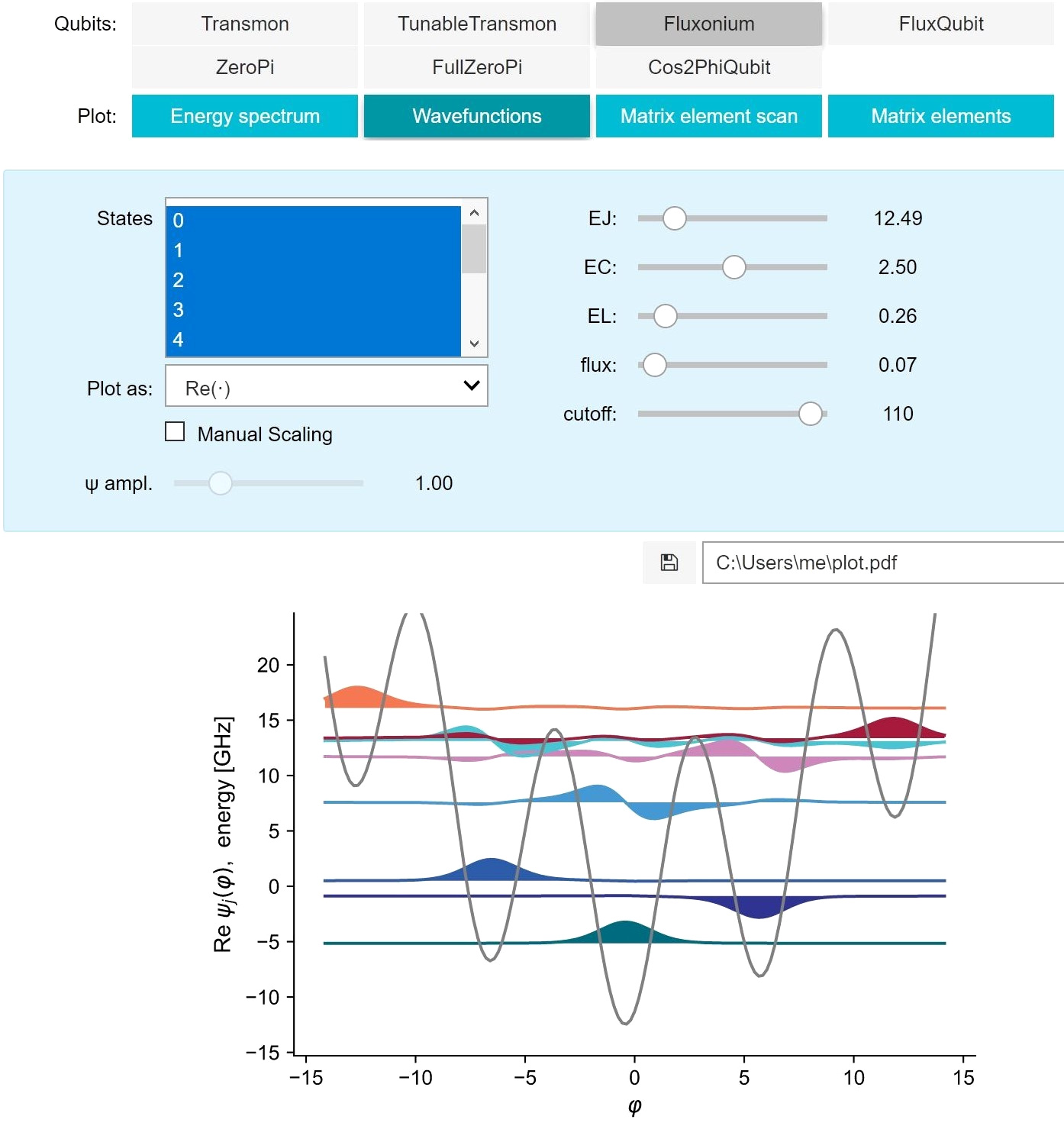}
	\caption{\scqubits graphical user interface for exploring properties of individual superconducting qubits.\label{fig:gui}}
	\end{figure*}

This section gives a broad overview and introduces the main building blocks of \scqubits based on concrete examples of their usage.
We stress that a jupyter notebook containing all of the source code in this manuscript can be found in the github examples repository (see Sec.~\ref{sec:onlinePresence}).
As a regular Python package, \scqubits is imported via
\begin{lstlisting}
import scqubits as scq
\end{lstlisting}
The simplest way to explore spectral properties of individual superconducting qubits is to invoke the dedicated graphical user interface via 
\begin{lstlisting}
scq.GUI()
\end{lstlisting}
which outputs the display shown in Fig.\ \ref{fig:gui}.
%\begin{widetext}
%\end{widetext}
With the exception of the initial call, usage of this interface requires  no further knowledge of Python and is particularly well-suited for beginners. In the following we describe the more powerful and flexible programmatic usage of \scqubits.

As an illustration of basic \scqubits functionality, we consider the transmon qubit. Like all qubit types implemented in \scqubits, \pyl{Transmon} and its flux-tunable variant \pyl{TunableTransmon} are realized as Python classes. Each class instance stores all relevant circuit parameters and control-field values, and provides a collection of methods used for common computations and visualization. (See Table \ref{table:qub1} for a summary of the most commonly used qubit class methods.) 
 An instance of the \pyl{TunableTransmon} class %, described by the charge-basis Hamiltonian
%\begin{equation}
%H= 4E_C n^2 -\frac{1}{2}E_J\sum_n\Big( |n\rangle\langle n+1| + |n+1\rangle\langle n|\Big)
%\end{equation}
is created by the following call which provides all necessary system parameters for initialization:
\begin{lstlisting}
tmon = scq.TunableTransmon(
    EJmax=30.0, 
    EC=1.2,
    d=0.01,
    flux=0.0,
    ng=0.0,
    ncut=30
)
\end{lstlisting}
The initialization arguments include the relevant circuit parameters: for a flux-tunable transmon, these are the maximum Josephson energy $E_{J\text{max}}$ from the SQUID loop,  charging energy $E_C$, and offset charge $n_g$ (further details are provided in appendix \ref{app:transmon}). If \scqubits is used inside a jupyter notebook, then 
\begin{lstlisting}
transmon = scq.TunableTransmon.create()
\end{lstlisting}
is an alternative way to create and initialize a \pyl{TunableTransmon} instance. The resulting graphical interface offers simple widgets to enter required parameters and displays the transmon circuit for reference. 
By default, energies are assumed to be given as frequencies in units of GHz, although this can be easily modified -- see Sec.~\ref{sec:units}. 

\begin{figure*}
	\centering
    \hspace*{-0.5cm}   
	\includegraphics[width=0.35\textwidth]{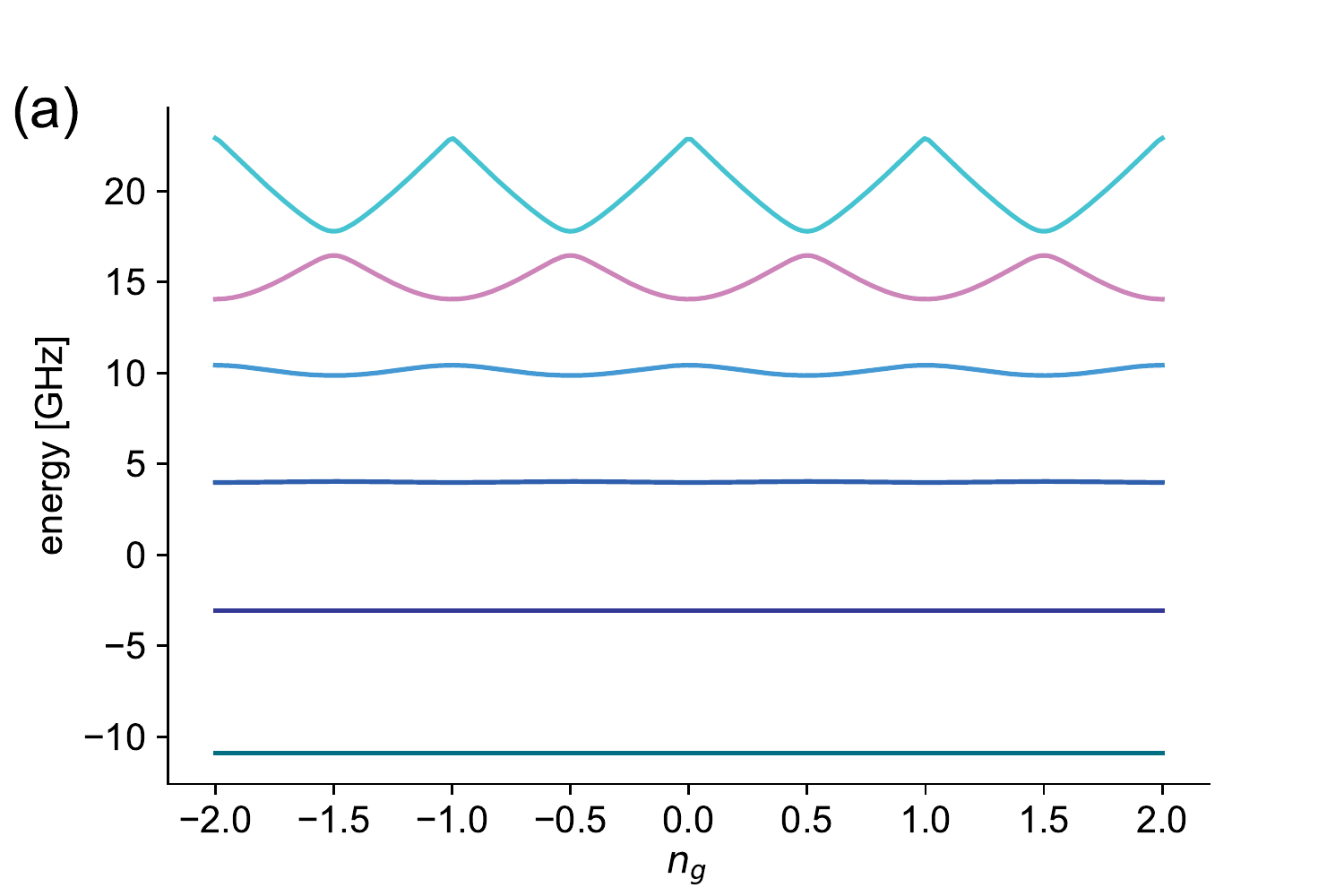}
    \hspace*{-0.5cm}   
	\includegraphics[width=0.35\textwidth]{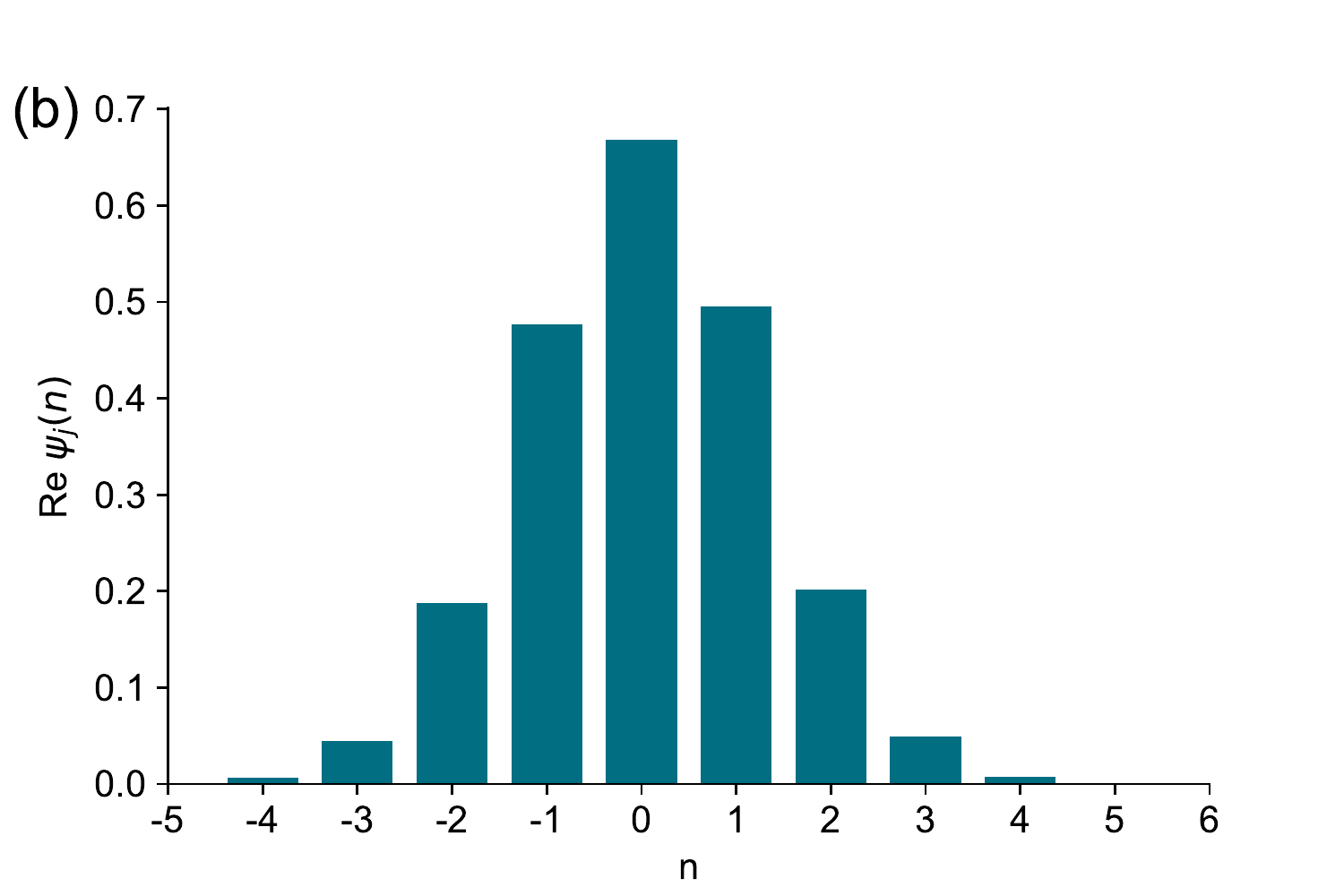}
    \hspace*{-0.5cm}   
	\includegraphics[width=0.35\textwidth]{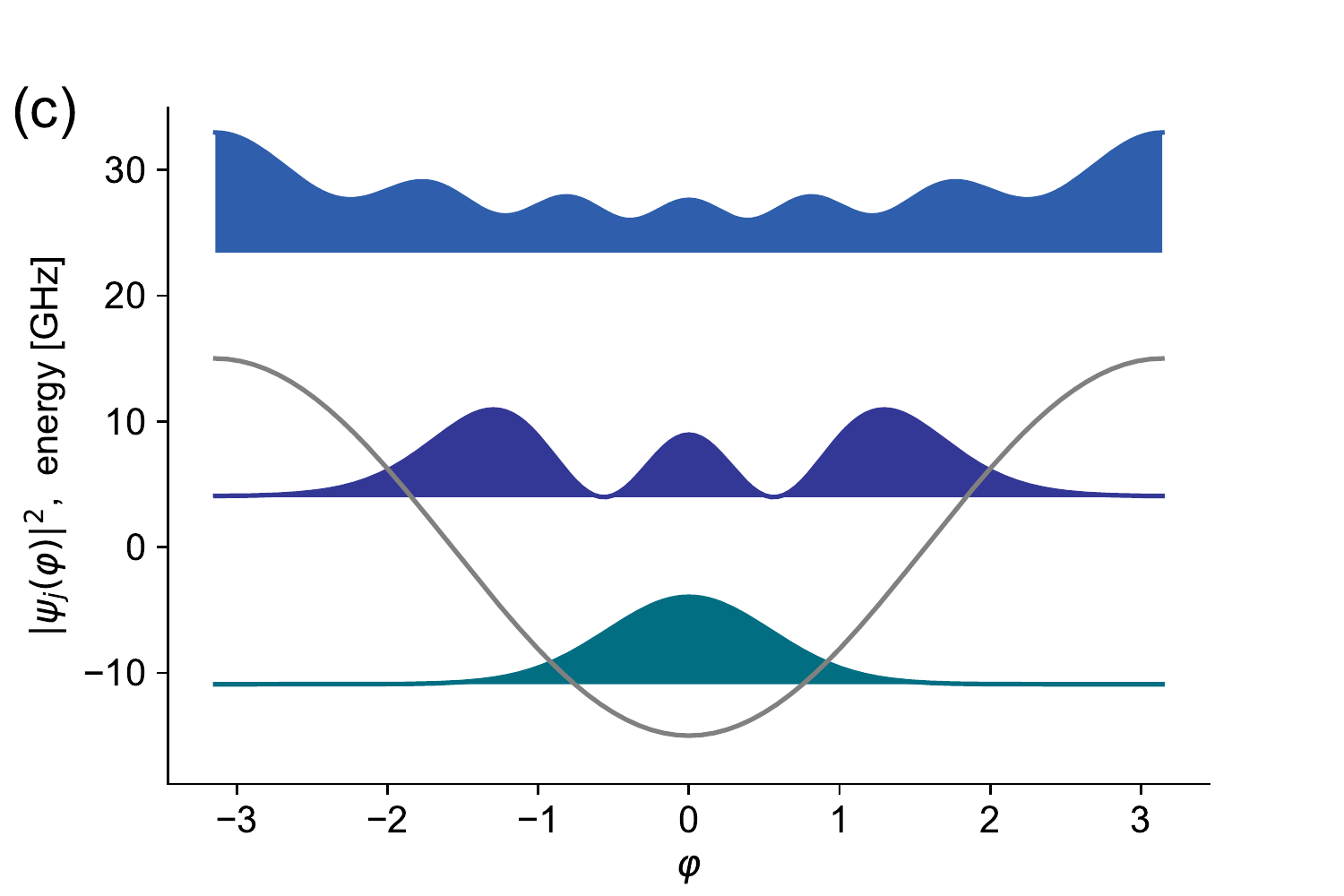}
    \caption{Visualization of transmon eigenenergies and eigenstates. (a) Plot of the lowest six transmon energy eigenvalues as a function of the offset charge $n_g$. (b) Plot of wavefunction amplitudes in the discrete charge basis, using \lstinline!plot_n_wavefunction()!. The slight asymmetry in the amplitudes with respect to $n=0$ originates from the choice of a nonzero offset charge $n_g$. (c) \lstinline!plot_phi_wavefunction()! graphs wavefunctions in the $\varphi$-basis, along with the underlying potential energy. Wavefunctions are offset vertically according to their eigenenergies.}
	\label{fig:fig1}
\end{figure*}

Finding eigenenergies and eigenstates of superconducting qubits invariably involves truncation of the infinite-dimensional Hilbert space.
In each qubit class, this can be controlled by setting the appropriate truncation parameter, such as \lstinline{ncut} for the transmon qubit.
While typical values are suggested in each widget, convergence with respect to this cutoff (and similar cutoffs in other qubit classes) must be established by the user (see \ref{sec:hilbertSpaceTrunc} below for a more detailed discussion).

\subsection{Computing and plotting energy spectra}
The energy eigenvalues of the transmon Hamiltonian are obtained by calling the \pyl{eigenvals} method. The optional parameter \pyl{evals_count} specifies the desired number of eigenenergies:
\begin{lstlisting}
tmon.eigenvals(evals_count=12)
\end{lstlisting}
Execution of this line yields a NumPy array of the lowest twelve energy eigenvalues. To plot eigenenergies as a function of one of the qubit parameters (here,  \pyl{EJmax},  \pyl{EC}, \pyl{flux}, or  \pyl{ng}), we generate an array of parameter values of interest, and then call the method  \pyl{plot_evals_vs_paramvals}. The latter takes as arguments the name of the parameter to be varied, an array of parameter values, and optionally the eigenvalue number. The following is an example for plotting the lowest six eigenenergies as a function of offset charge \pyl{ng} for 220 equally spaced points in the interval $n_g\in[-2,2]$:
\begin{lstlisting}
ng_list = np.linspace(-2, 2, 220)
tmon.plot_evals_vs_paramvals('ng', ng_list,
                             evals_count=6)
\end{lstlisting}
Figure \ref{fig:fig1}(a) shows the resulting output. Plotting routines in \scqubits rely on matplotlib, and generally return a tuple of a \pyl{Figure} and an \pyl{Axes} object to enable post-processing of the plot, if desired.

The full eigensystem consisting of both eigenvalues and eigenvectors is obtained through the method \pyl{eigensys}:
\begin{lstlisting}
evals, evecs = tmon.eigensys()
\end{lstlisting}
For the transmon qubit, this calculation is based on\footnote{In some cases (e.g., $0-\pi$ or the $\cos2\phi$ qubit), \scqubits, uses \pyl{scipy.sparse.linalg.eigsh} for diagonalization.} \pyl{scipy.linalg.eigh} which performs diagonalization of the Hamiltonian matrix expressed in the charge basis. As dictated by the SciPy package, the eigenvector corresponding to the \pyl{j}-th eigenvalue is the Numpy array \pyl{evecs.T[j]}.

\subsubsection{Hilbert space truncation and convergence}
\label{sec:hilbertSpaceTrunc}
    As mentioned above, obtaining qubit spectra requires truncating the Hilbert space to some finite dimension.
    Users can control the Hilbert space dimension used by \scqubits, by passing appropriate values for \pyl{ncut} and/or \pyl{grid} parameters to qubit class constructors \footnote{In some qubits with multiple degrees of freedom, these variable names may be slightly modified to reflect which degrees of freedom are being addressed. See the API documentation \cite{APIdocs} as well as the Appendices for details. }. These parameters effectively define the number of basis states that are used during diagonalization. In qubits with multiple degrees of freedom, users need to set a cutoff independently for each one.
    
    Choosing cutoffs or grid sizes that are too small may lead to inaccurate results. 
    The specific parameter values required for convergence naturally depend on the qubit type, circuit energies, as well as how many eigenenergies and/or eigenvectors the user wishes to obtain. While in some simple cases (e.g., transmon qubit - see \cref{app:transmon}) one can establish stringent cutoff requirements, for most circuits convergence must be established by trial and error.
    
    A heuristic approach to this end is to repeat calculations with successive increases in cutoffs and grid sizes until results are essentially unchanged (within the desired accuracy). Once this is achieved, errors are typically limited by the default tolerances set by the \pyl{scipy} routines which \scqubits internally uses for matrix diagonalization.

\subsection{Plotting wavefunctions}
The transmon qubit is a circuit with a single degree of freedom. Hence, its wavefunctions are readily plotted in the two bases natural for the transmon: the discrete charge basis $\psi_j(n)=\langle n |\psi_j\rangle$, and the continuous phase basis $\psi_j(\varphi) = \langle \varphi | \psi_j\rangle$. The \pyl{TunableTransmon} class offers two methods for this purpose:
\begin{lstlisting}
tmon.plot_n_wavefunction(which=0, mode='real');
tmon.plot_phi_wavefunction(which=[0,2,6],
                           mode='abs_sqr');
\end{lstlisting}
Figure \ref{fig:fig1}(b-c) shows the generated graphs. Above, the keyword argument
\pyl{which} specifies the selection of wavefunctions to plot. In the first case, \pyl{which=0} indicates that only the ground state wavefunction (index 0) should be displayed. In the second case, \pyl{which} is assigned a list consisting of multiple such indices, resulting in a plot showing several wavefunctions at once. The \pyl{mode} option determines how to plot the wavefunction which is generally complex-valued. The self-explanatory options are \pyl{'real'}, \pyl{'imag'}, \pyl{'abs'}, and \pyl{'abs_sqr'}.   

%\begin{center}
%	\small
%	\rowcolors{1}{gray!10}{white}
%	\arrayrulecolor{deepblue} 
%	\setlength{\tabcolsep}{0pt}
%	\setlength\extrarowheight{5pt}
%	\begin{tabular}{p{0.3\columnwidth} >{\raggedright\arraybackslash}p{0.7\columnwidth}}
%		\rowcolor{lightblue}
%		\textbf{\pyl{mode}} & \textbf{description}   \\\toprule
%		\pyl{'real'}        & real part of amplitudes           \\
%		\pyl{'imag'}        & imaginary part of amplitudes      \\
%		\pyl{'abs'}         & absolute value of amplitudes      \\
%		\pyl{'abs_sqr'}     & abs.\ value squared of amplitudes 
%	\end{tabular}
%\end{center}

For one-dimensional wavefunctions in $\varphi$ basis (``position'' basis), the plot mimics the textbook format widely used for 1d spectra: wavefunctions of the eigenstates are shown alongside the potential-energy function, and are offset vertically by their corresponding eigenenergies. 

\subsection{Evaluating and visualizing matrix elements}
\begin{figure}
	\centering
    \hspace*{-1cm}   
	\includegraphics[width=1.2\columnwidth]{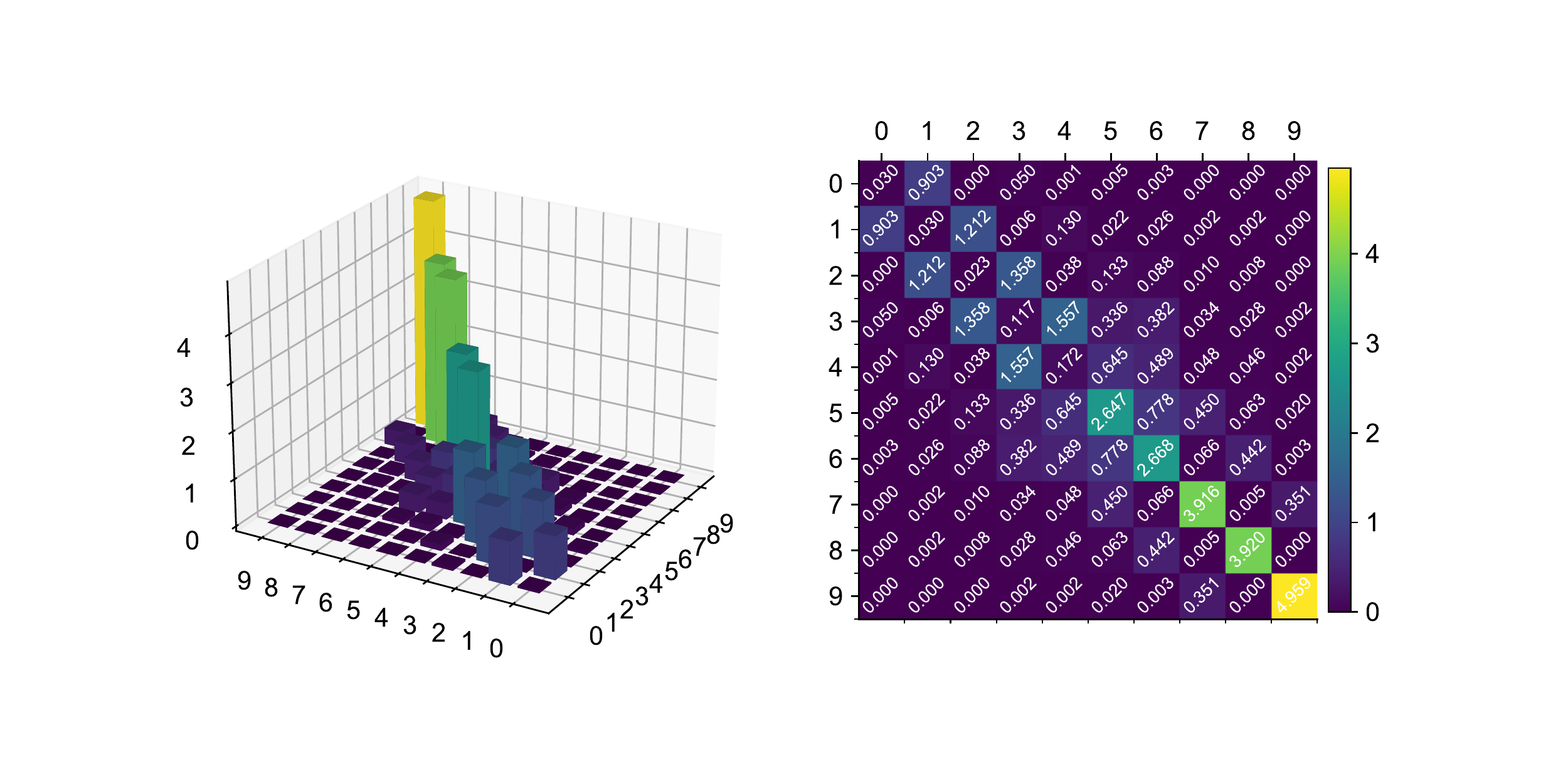}
    \vspace{-1cm}
    \caption{Visualization of the matrix elements of the transmon charge operator $\langle \psi_i | \hat{n} | \psi_j\rangle$, evaluated with respect to the transmon eigenstates using \texttt{plot\_matrixelements}. Numerical values can be included with the option \pyl{show\_numbers=True}. }
	\label{fig:figure2}
\end{figure}

Matrix elements of qubit operators play an important role in determining coupling strengths between a qubit and another quantum system. They are also critical for the sensitivity of the qubit to various sources of noise. In the case of the transmon qubit, for example,  matrix elements of the charge operator are of frequent interest. This charge operator, $\hat{n}$, is accessible through the class method \pyl{n_operator}. Evaluation of the matrix elements for the current set of parameters is implemented by  the method \pyl{matrixelement_table}.  An overview plot of  matrix elements with respect to the lowest ten transmon eigenstates is obtained by
\begin{lstlisting}
tmon.plot_matrixelements('n_operator',
                         evals_count=10,
                         show_numbers=True)
\end{lstlisting}
where the reference to the operator is provided in string format.
The resulting plot is shown in Fig.~\ref{fig:figure2}.

% The resulting NumPy array can be displayed in an alternative way including information about the individual values by using the option \pyl{show\_numbers=True}:
% \begin{lstlisting}
% tmon.plot_matrixelements('n_operator', 
%                          show_numbers=True,
%                          show3d=False,
%                          evals_count=10)
% \end{lstlisting}
% This produces the plot presented in Fig.\ \ref{fig:figure3}(a).

\begin{figure}
	\centering
    \includegraphics[width=1.1\columnwidth]{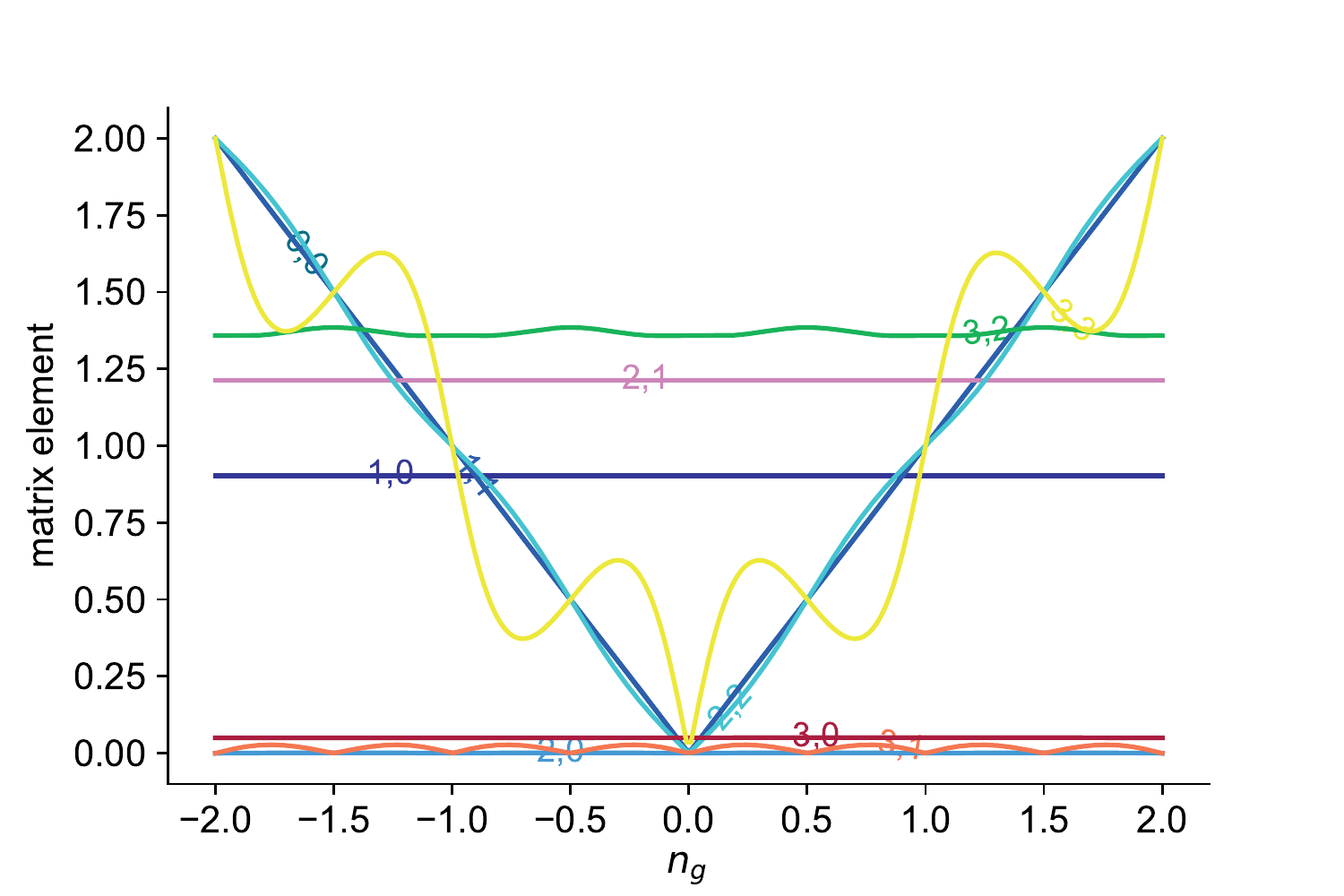}
\caption{Plot of select matrix elements as a function of the external offset-charge parameter $n_g$.
}
	\label{fig:figure3}
\end{figure}

Occasionally, it is further useful to plot matrix elements as a function of an external parameter. This is accomplished by  \pyl{plot_matelem_vs_paramvals}  Applied to the transmon charge matrix elements, we can easily visualize the dependence on the offset charge $n_g$:
\begin{lstlisting}
ng_list = np.linspace(-2, 2, 220)
tmon.plot_matelem_vs_paramvals('n_operator',
                               'ng', ng_list,
                               select_elems=4)
\end{lstlisting}
Here, the names of both the operator and the external parameter are given as string arguments, followed by the array of values for the external parameter. Finally, \pyl{select_elems=4} specifies that all matrix elements $\langle \psi_i | n |\psi_j\rangle$ with $0\le i,j \le 3$ are requested in the plot. The output is shown in Fig.\ \ref{fig:figure3}.

\subsection{Units}
\label{sec:units}

\scqubits allows the user to associate specific units with the energy values they provide when instantiating various qubit classes. The knowledge by \scqubits of implied units is necessary for calculations that involve coherence time estimations (see Sec.~\ref{sec:coherence}), but also come in handy for automatic labeling of plot axes. 
All energies are assumed to be expressed in terms of frequencies (not angular frequencies), with the default being GHz. 
Other supported units are MHz, kHz and Hz. A list containing these possible choices can be shown with the \pyl{show_supported_units} function.
The currently set units can be obtained with the \pyl{get_units} function, and changed with \pyl{set_units}, like so:
\begin{lstlisting}
scq.get_units()
scq.set_units('MHz')
\end{lstlisting}
\scqubits also includes several helper functions for convenient conversion between the currently set units and Hz, which include \pyl{to_standard_units}, \pyl{from_standard_units} and  \pyl{units_scale_factor}.

\section{Composite Hilbert spaces and interface with QuTiP}
\label{sec:compositeHilberSpaces}

An important aspect of modeling superconducting circuits is the ability to study composite systems. \scqubits provides an easy mechanism to explore setups that may consist of multiple qubits as well as harmonic (and weakly anharmonic) oscillators. 
Along with providing an easy way of constructing composite-system Hilbert spaces, and calculating and visualizing their many properties, \scqubits also allows for easy exporting of effective Hamiltonians to QuTiP, an established toolbox for studying stationary and dynamical properties of closed and open quantum systems. 
At the heart of this functionality is the \pyl{HilbertSpace} class, which provides the data structures and methods for handling composite Hilbert space objects, and which we briefly explore in the sections below. 

\subsection{Example system: two transmons coupled to a harmonic mode}
Transmon qubits can be capacitively coupled to a common harmonic mode, realized by an LC oscillator or a transmission-line resonator. The Hamiltonian describing such a composite system is given by
\begin{align}
    \Hop&= E_\text{osc} \aop \dg \aop + \sum_{j=1,2} \Hop_\text{tmon,j} + \sum_{j=1,2} g_j \noper_j(\aop+\aop \dg),
\end{align}
where $j=1,2$ enumerates the two transmon qubits, $E_\text{osc}$ is the single-photon energy for the resonator. Furthermore, $\noper_j$ is the charge number operator for qubit $j$, and $g_j$ is the coupling strength between qubit $j$ and the resonator.

The first step consists of creating the objects describing the individual building blocks of the full Hilbert space. Here, these will be the two transmons and one oscillator:
\begin{lstlisting}
tmon1 = scq.Transmon(
    EJ=40.0,
    EC=0.2,
    ng=0.3,
    ncut=40,
    truncated_dim=4   
)

tmon2 = scq.Transmon(
    EJ=15.0,
    EC=0.15,
    ng=0.0,
    ncut=30,
    truncated_dim=4
)

resonator = scq.Oscillator(
    E_osc=4.5,
    truncated_dim=4 
)
\end{lstlisting}
The significance of \pyl{truncated_dim} lies in a simple hierarchical diagonalization scheme. Specifically, each subsystem is diagonalized separately in step one. Subsequently, the lowest few bare subsystem eigenstates up to truncation level \pyl{truncated_dim} are fed forward into \pyl{HilbertSpace}.

\subsection{Creating the \pyl{HilbertSpace} object}
The desired \pyl{HilbertSpace} object can be created in two ways: either by utilizing the GUI displayed via \pyl{hs = scq.HilbertSpace.create()}, or programmatically by initializing the \pyl{HilbertSpace} object with a list of all subsystems and then specifying individual interaction terms:
\begin{lstlisting}
hs = scq.HilbertSpace([tmon1,tmon2,resonator])
\end{lstlisting}

\subsection{Specifying interactions}
Interaction terms describing the coupling between subsystems (or modifying a single subsystem itself) can be specified via the method \pyl{add_interaction} in three different ways.

\paragraph{1.\ Operator-product based interface:}
Interaction terms involving multiple subsystems $S=1,2,3,\ldots$ are often of the form
\begin{align*}
\Vop &= g\, \Aop_1 \Aop_2 \Aop_3 \cdots \qquad \text{or}\\
\Vop &= g\, \Bop_1 \Bop_2 \Bop_3 \cdots \,+ g^* (\Bop_1 \Bop_2 \Bop_3 \cdots)^\dag
\end{align*}
where the operators $\Aop_j$, $\Bop_j$ act on subsystem $j$. (In the first case, the operators $\Aop_j$ are expected to be hermitian.)
This structure is captured in the following way:
\begin{lstlisting}
# coupling resonator-tmon1
g1 = 0.1  
operator1 = tmon1.n_operator()
operator2 = resonator.creation_operator() +           
            resonator.annihilation_operator()

hs.add_interaction(
    g=g1,
    op1=(operator1, tmon1),  # (matrix, subsys)
    op2=(operator2, resonator)
)

# coupling resonator-tmon2
g2 = 0.2  
hs.add_interaction(
    g=g2,
    op1=tmon2.n_operator,    # class method
    op2=resonator.creation_operator,
    add_hc=True
)
\end{lstlisting}
In this operator-product interface, \pyl{op1}, \pyl{op2},$\ldots$ are either of the form \pyl{(<array>, <subsystem>)}, i.e., a tuple with an array or sparse matrix in the first position, and the corresponding subsystem in the second position; or of the form \pyl{<callable>}, i.e., the operator is provided as a callable method which will automatically yield the subsystem the operator function is bound to. (These two choices can be mixed and matched.) The option \pyl{add_hc=True} adds the hermitian conjugate to the specified interaction term.

Note that interactions based on only one operator are possible (simply drop all but the \pyl{op1} entry). One example use case of this is the creation of a higher-order non-linearity $a^\dagger a^\dagger a^\dagger a a a$ in a Kerr oscillator.

\paragraph{2.\ String-based interface: }
The \pyl{add_interaction} method can also be used to define the interaction in string form, by providing an expression that can be evaluated by the Python interpreter.
\begin{lstlisting}
hs = scq.HilbertSpace([tmon1, tmon2, resonator])
g3 = 0.1

hs.add_interaction(
    expr="g3 * cos(n) * adag",
    op1=("n", tmon1.n_operator(), tmon1),
    op2=("adag", resonator.creation_operator),
    add_hc=True
)
\end{lstlisting}
Here, \pyl{expr} is a string used to define the interaction as a Python expression. It may use variables that are already defined globally, and operators given by the names provided in \pyl{op1}, \pyl{op2}, $\ldots$. 

\paragraph{3.\ \pyl{Qobj} interface: }
Finally, \pyl{add_interaction} can be used to directly add a QuTiP \pyl{Qobj} that has already been properly wrapped with identities:
\begin{lstlisting}
import qutip as qt

# Generate a Qobj
g = 0.1
a = qt.destroy(4)
kerr = a.dag() * a.dag() * a * a
id = qt.qeye(4)
V = g * qt.tensor(id, id, kerr)

hs.add_interaction(qobj=V)
\end{lstlisting}

\subsection{Obtaining the Hamiltonian and spectrum}
With the interactions specified, the full Hamiltonian of the coupled system can be obtained via the method \pyl{hamiltonian},
\begin{lstlisting}
dressed_hamiltonian = hs.hamiltonian()
\end{lstlisting}
which represents $H$ in the basis of the bare product states composed of subsystem eigenstates.
Since the Hamiltonian obtained this way is a proper \pyl{Qobj}, it can easily be handed over to QuTiP's time evolution routines such as \pyl{mesolve}. Eigenenergies and eigenstates can now either be obtained  via the usual \scqubits methods, 
\begin{lstlisting}
evals = hs.eigenvals()  # or
evals, evecs = hs.eigensys()
\end{lstlisting}
or by invoking QuTiP methods on the Hamiltonian itself, e.g., \pyl{hs.hamiltonian.eigenstates()}.

\section{Sweeping over external parameters}
\label{sec:sweeps}
Determining the dependence of physical observables on one or multiple external parameter(s) is a common way to gain intuition for the properties and behavior of a system. Such parameter sweeps can be performed with \scqubits on multiple levels: 
(1) at the level of a single qubit, and (2) at the level of a composite quantum system.

At the single-qubit level, each qubit class provides several methods that enable producing parameter sweep data and plots. Central quantities of interest, in this case, are energy eigenvalues and matrix elements -- in particular, their dependence on parameters like flux or offset charge.
The relevant methods available for every implemented qubit class are:
\begin{itemize}
 \item \pyl{get_spectrum_vs_paramvals}   \\
 sweep eigenvalues and eigenvectors  
 \item \pyl{get_matelements_vs_paramvals} \\
 sweep matrix elements
 \item \pyl{plot_evals_vs_paramvals}      \\
 plot eigenenergy sweep 
\item \pyl{plot_matelem_vs_paramvals}    \\plot matrix element sweep
\end{itemize}
                                      
\subsection{Creating a \pyl{ParameterSweep} object}                     
Composite Hilbert spaces, as implemented by \pyl{HilbertSpace} objects, are naturally richer than individual qubits. A variety of parameter sweeps can be considered, including multi-dimensional sweeps over a collection of different parameters. 
For flexible parameter scans, \scqubits provides the \pyl{ParameterSweep} class. To illustrate its usage, we reuse a composite Hilbert space akin to the one presented above: two tunable transmon qubits capacitively coupled to an oscillator.

The \pyl{ParameterSweep} class facilitates computation of spectra as function of one or multiple external parameter(s). For efficiency in computing a variety of derived quantities and creating plots, the computed bare and dressed spectral data are stored internally. 
A \pyl{ParameterSweep} object is initialized by providing the following parameters:
\begin{enumerate}
\item \pyl{hilbertspace}: a \pyl{HilbertSpace} object that describes the quantum system of interest
\item  \pyl{paramvals_by_name}: a dictionary that maps each parameter name (string) to an array of parameter values 
\item  \pyl{update_hilbertspace}: a function that defines how parameter changes affect the system
\item  \pyl{subsys_update_info}: (optional) for potential speed-up, specify which subsystems undergo changes as each of the parameters is varied
\item \pyl{deepcopy}: (optional) determines whether the HilbertSpace object and all constituents should be duplicated and disconnected from the global objects
\item  \pyl{num_cpus}: (optional) number of CPU cores requested for the sweep evaluation
\end{enumerate}

These ingredients all enter as initialization arguments of the \pyl{ParameterSweep} object. Once initialized, spectral data is generated and stored.

In our example, we consider the strength of a global magnetic field as the parameter to be changed. This field determines the magnetic fluxes for both qubits, in proportions according to their SQUID loop areas. We will reference the flux for transmon 1, and express the flux for transmon 2 in terms of it via an area ratio. In addition, we will vary the offset charge of transmon 2.

The following code illustrates this functionality:
\begin{lstlisting}
# combine parameter names and values 
# in a dictionary
paramvals_by_name = {
    "flux": np.linspace(0.0, 2.0, 171), 
    "ng": np.linspace(-0.5, 0.5, 49)
}

area_ratio = 1.2
def update_hilbertspace(flux, ng):  
    # function that defines how Hilbert space 
    # components are updated
    tmon1.flux = flux
    tmon2.flux = area_ratio * flux
    tmon2.ng = ng

# dictionary specifying which subsystems are 
# affected by changing each parameters   
subsys_update_info = {"flux": [tmon1, tmon2],
                      "ng": [tmon2]}

# create the ParameterSweep object
sweep = scq.ParameterSweep(
    hilbertspace=hs,
    paramvals_by_name=paramvals_by_name,
    update_hilbertspace=update_hilbertspace,
    evals_count=20,
    subsys_update_info=subsys_update_info,
    num_cpus=4
)
\end{lstlisting}

In the  code above, \pyl{update_hilbertspace} directly manipulates transmon instances via their global references. Alternatively, \pyl{HilbertSpace} constituents can be accessed via \pyl{sweep.hilbertspace[<id_str>]} where \pyl{id_str} is a string identifier either provided explicitly at initialization of object instances, or autogenerated by \scqubits. (Details and examples of this functionality are available in the documentation and example notebooks -- see Sec.~\ref{sec:onlinePresence}.)

\subsection{Generated \pyl{ParameterSweep} data}
Much of the computed data that is stored and immediately retrievable after this sweep. The stored data is accessed as if \pyl{ParameterSweep} were a Python \pyl{dict} with the following string keys:
\begin{enumerate}
    \item \pyl{"evals"}, \pyl{"evecs"}:\\ dressed eigenenergies and eigenstates
    \item \pyl{"bare_evals"}, \pyl{"bare_evecs"}:\\ bare eigenenergies and eigenstates for each subsystem
    \item \pyl{"lamb"}, \pyl{"chi"}, \pyl{"kerr"}:\\ dispersive energy coefficients 
\end{enumerate}

Data are returned as a \pyl{NamedSlotsNdarray}, a subclass of  the regular NumPy \pyl{ndarray}. It features several convenient new slicing options, such as slicing by parameter name, or slicing by reference to a particular parameter value. See \cite{onlineDocs} for a more more detailed discussion as well as examples. 

\subsection{Transition plots}
Energy spectra obtained in single-tone or two-tone spectroscopy always represent transition energies, rather than absolute energies of individual eigenstates. To generate data mimicking this, appropriate differences between eigenenergies must be taken.

The methods for generating transition energy data and plotting them are \pyl{transitions} and \pyl{plot_transitions}. To create a plot, we first ``pre-slice''  the \pyl{ParameterSweep} instance which specifies a sweep along a single axis. Then, the \pyl{plot_transitions} method can be called\footnote{A notebook with exact parameters used to generate this plot is included in the scqubits-examples github repository -- see Sec.~\ref{sec:onlinePresence}.}:
\begin{lstlisting}
sweep["ng":0.0].plot_transitions();
\end{lstlisting}
\vspace{-0.2cm}
\includegraphics[width=1.0\columnwidth]{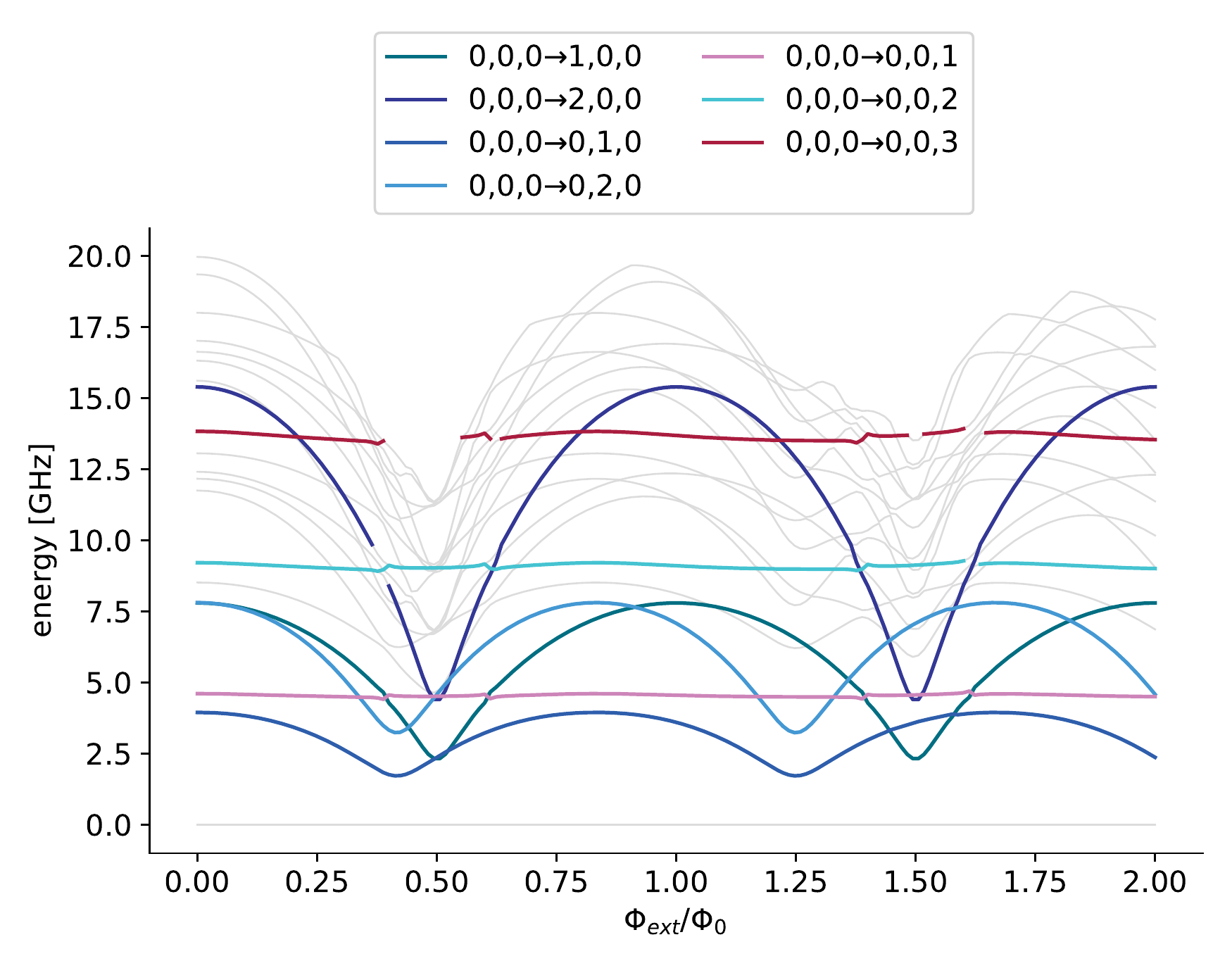}\\ 
\vspace{-0.5cm}

When coloring transitions according to the dispersive-limit product state labels, one potential artifact is nearly unavoidable: whenever states undergo avoided crossings and the dispersive limit breaks down, coloring must discontinuously switch from one branch to another. \scqubits attempts to interrupt coloring in such regions. However, if the avoided crossing occurs over a range comparable to the parameter value spacing, then discontinuities from connecting separate branches will remain visible.

\paragraph{Transition plot options: }
The generated transition plot above is based on a number of default settings, including: (i) the origin of each transition is the system's ground state, (ii) single-photon transitions are plotted in light grey and (iii) transitions within each individual subsystem are marked separately in color and accompanied by a legend. This is possible in regions where the dispersive approximation holds, i.e., hybridization between subsystems remains weak. Labels in the legend are excitation levels of individual systems: ((0,0,0), (1,0,0)) denotes a transition from the ground state to the state with subsystem 1 in the first excited state, and subsystems 2 and 3 in their respective ground states.

Many aspects of transition energy plots can be changed. The following  illustrates a subset of options that change which transitions are plotted and how.
\begin{description}
\lstitem[morekeywords={coloring:}]{coloring:} Coloring based on dispersive-transition identification can be switched off by setting \pyl{coloring="plain"}. 
\lstitem[morekeywords={subsystems:}]{subsystems:}
By default, dispersive-transition coloring includes all subsystems. If only transitions for a single or smaller set of subsystem(s) should be highlighted, then these can be specified in list form, e.g.\ \pyl{subsystems=[tmon1]}.
%leads to wrong coloring
%\lstitem[morekeywords={initial,final:}]{initial,final:}
\lstitem[morekeywords={}]{initial,final:}
The ground state is the default origin for all transitions. In case of thermal excitations, other states can be of interest as initial states. Specification of an alternative initial state uses dispersive labeling of states, e.g., \pyl{initial=(1,0,0)} uses the 1st excited of the first subsystem as the initial state.
\lstitem[morekeywords={photon_number:}]{photon_number:}
To model $n$-photon transitions, setting \pyl{photon_number=}$n$ yields transition energies divided by $n$.
\lstitem[morekeywords={sidebands:}]{sidebands:}
For \pyl{sidebands=True} sideband transitions with multiple subsystems changing excitation levels are included in color highlighting and legend.
\end{description}

\subsection{Custom sweep data}
\pyl{ParameterSweep} automatically generates data commonly needed in studying a multi-component quantum system. Other quantities of interest can be generated by defining a custom sweep function,
\begin{lstlisting}
def custom_func(paramsweep, paramindex_tuple, 
                paramvals_tuple, **kwargs):
    ...
    return data
\end{lstlisting}
This function returns the data to be calculated for each parameter choice. The custom sweep is then performed upon calling
\begin{lstlisting}
<ParameterSweep>.add_sweep(custom_func, 
                           "custom name")
\end{lstlisting}
The computed data is subsequently accessible via 
\begin{lstlisting}
<ParameterSweep>["custom name"]
\end{lstlisting}

\section{Estimation of coherence times}
\label{sec:coherence}

\scqubits provides extensive functionality that allows users to estimate coherence times of various qubits. 
Very general methods \pyl{t1} and \pyl{tphi} are implemented for each noisy qubit, which can be used for calculations of depolarization as well as pure dephasing times for almost arbitrary noise processes (see \cref{sec:depolariztion,sec:dephasing} for more details). 
Furthermore, a large variety of predefined methods corresponding to more specific noise channels (e.g., dephasing due to $1/f$ charge noise, or depolarization from coupling to a transmission line) are implemented as well (for a list, see Table \ref{table:predefNoiseChannels}). 

Due to different qubit properties and circuit topologies, each qubit is affected by some specific subset of these predefined noisy processes. 
To see which channels are currently implemented for any given qubit, one can run the command (here shown for the case of a transmon)
\begin{lstlisting}
tmon.supported_noise_channels()
\end{lstlisting}
\lstset{style=output,inputpath=code}
\vspace*{-3mm}
\begin{lstlisting}
['tphi_1_over_f_flux',
 'tphi_1_over_f_cc',
 'tphi_1_over_f_ng',
 't1_capacitive',
 't1_flux_bias_line',
 't1_charge_impedance']
\end{lstlisting}
\lstset{style=codeblock,inputpath=code,}
This returned list contains methods with self-explanatory names that either start with \pyl{t1} or \pyl{tphi} and represent the depolarization or pure dephasing processes, respectively. 
Each of these methods can then be called directly to obtain the corresponding coherence time (which will take into account the current units setting -- see \cref{sec:units}). 
For example, in order to calculate the pure dephasing time due to $1/f$ flux noise, one can simply execute
\begin{lstlisting}
tmon.tphi_1_over_f_flux()
\end{lstlisting}
Each available method can be provided with a set of custom parameters that give the user a means to fine-tune various noise process properties (e.g., bath temperature, $1/f$ flux-noise strength, etc.). There are also common method options that specify which qubit levels should be considered (levels 0 and 1 are assumed by default), whether a rate instead of a time should be returned, or whether depolarization rates should include both upwards and downwards transitions.
A more complicated method call could therefore take the form 
\begin{lstlisting}
tmon.t1_charge_impedance(i=3, j=1, 
                         Z=50, 
                         T=0.100, 
                         get_rate=True, 
                         total=False)
\end{lstlisting}
This returns a depolarization rate (not time) between levels 3 and 1, due to the qubit coupling to a 50$\,\Omega$ transmission line, at a temperature of $100\,$mK. In this case, the impedance (parameter \pyl{Z}) can be a constant, or alternatively an angular frequency-dependent Python function that will be evaluated at the frequency difference between levels $i$ and $j$. 

In the next sections we briefly outline the basic  physics and assumptions underlying the estimation of the various pure dephasing and depolarization times.  

\subsection{Dephasing due to $1/f$ noise}
\label{sec:dephasing}
Dephasing noise leads to loss of coherence, i.e., the relative phases relevant in superpositions of multiple states are lost over time.
One of the most important kinds of noise affecting superconducting qubits is $1/f$ noise, which leads to slow fluctuations of the energy-level spacing.
The spectral density function characterizing this noise is given by
\begin{align}
   S(\omega) &=  \frac{2 \pi A_{\lambda}^{2} }{|\omega|}.
    \label{eq:specDens1overf}
\end{align}
%with $\lambda$ representing a noisy circuit parameter such as flux or charge for example. 
Here, $A_{\lambda}$ corresponds to the amplitude or strength of the particular noise channel $\lambda$, such as charge or flux. \scqubits uses sensible default values for this quantity based on the literature. Alternative values, however, can be set when provided by the user.
The pure dephasing time due to a noise channel labeled $\lambda$ (away from sweet spots\footnote{Currently \scqubits returns a value of \pyl{numpy.inf} at sweet spots. Higher order corrections will be added at a later time. }) is given by \cite{Ithier05,Koch2007a} 
\begin{align}
   T_{\phi} &=  A_{\lambda} \frac{\partial \omega_{01}}{\partial \lambda}  \sqrt{ 2| \ln \omega_{\rm low} t_{\rm exp} |}
    \label{eq:tphi1overf}
\end{align}
with $t_{\rm exp}$ representing the measurement time, and $\omega_{\rm low}$ the low-frequency cutoff. (If not provided by the user then defaults of $t_{\rm exp}=10\,\mu$s and  $\omega_{\rm low}/2\pi=1\,$Hz are used.) 

As already hinted above, some qubits provide predefined methods for estimating effects due to $1/f$ flux, charge as well as Josephson junction critical-current noise channels, named \pyl{tphi_1_over_f_flux}, \pyl{tphi_1_over_f_charge} and \pyl{tphi_1_over_f_cc} respectively. Each qubit also implements a more general method \pyl{tphi_1_over_f} which accepts a user-defined operator $\partial_{\lambda} \Hop$ (for an arbitrary, user-defined noise channel $\lambda$), that is then internally used by \scqubits to implement Eq.~\ref{eq:tphi1overf}. 
%See \cite{APIdocs} for more information. 

\subsection{Depolarization}
\label{sec:depolariztion}

Noise may also cause depolarization of the qubit by inducing spontaneous transitions among eigenstates. \scqubits uses the standard perturbative approach (Fermi's Golden Rule) to approximate the resulting transition rates due to different noise channels.
The rate of a transition from state $i$ to state $j$ can be expressed as
\begin{align}
   \Gamma_{ij} &=  \frac{1}{\hbar^2} |\langle i| \Bop_{\lambda} |j \rangle|^2 S(\omega_{ij}),
    \label{eq:t1general}
\end{align}
where $\Bop_\lambda$ is the noise operator, and $S(\omega_{ij})$ the spectral density function evaluated at the angular frequency associated with the transition frequency, $\omega_{ij} = \omega_{j} - \omega_{i}$.
$\omega_{ij}$, which is positive in the case of  decay (the qubit emits energy to the bath), and negative in case of excitations (the qubit absorbs energy from the bath).
Unless stated otherwise (see channel-specific documentation \cite{onlineDocs}), it is assumed that the depolarizing noise channels satisfy detailed balance, which implies
\begin{align}
    \frac{S(\omega)}{S(-\omega)} &=  \exp \left(\frac{\hbar \omega}{k_B T}\right),
    \label{eq:specDens2}
\end{align}
where $T$ is the bath temperature, and $k_B$ Boltzmann's constant.
By default, all \pyl{t1} methods estimate the  depolarization times based on the \emph{sum of the upward and downward rates}. This behavior is controlled by the argument \pyl{total}, which can be modified by the user. For example, setting \pyl{total=False} will calculate only a single-directional transition rate from the state indexed $i$ to the state indexed $j$.

As in the case of $1/f$ dephasing, each qubit implements a subset of predefined depolarization methods such as \pyl{t1_capacitive} or \pyl{t1_flux_bias_line} for example (see \cite{onlineDocs} for details on what methods are implemented for each qubit). Coherence times due to arbitrary depolarization processes can also be readily calculated. This is done using a general method \pyl{t1}, which accepts an arbitrary, user-constructed $\Bop_{\lambda}$ operator, as well as a custom spectral density function $S(\omega)$.

%\begin{widetext}
\begin{table*}[t]
     \tabcaption{List of predefined methods for estimating coherence times due to depolarization ($T_{1}$) or pure dephasing ($T_{\phi}$). Different subsets of these are implemented for each qubit class. (See the API documentation \cite{APIdocs} for how to change default behavior.)}
	\small
	%\rowcolors{1}{gray!10}{white}
	\arrayrulecolor{deepblue} 
	\setlength{\tabcolsep}{0pt}
	\setlength\extrarowheight{5pt}
	\begin{tabular}{p{0.33\textwidth} >{\raggedright\arraybackslash}p{0.67\textwidth}}
		\rowcolor{lightblue}
		\textbf{Method name} & \textbf{Description}     \\\hline
		\rowcolor{gray!10}
        \scshape $T_{1}$ depolarization processes   & \\
        \pyl{t1_capacitive}  & Capacitive loss due to dielectric dissipation \cite{Pop2014,Smith2020} \\ 
        \pyl{t1_charge_impedance} & Loss due to charge-coupling to an impedance (e.g., open transmission line) \cite{Schoelkopf2002a,Ithier2005}  \\ 
        \pyl{t1_flux_bias_line}  & Loss due to current fluctuations in the flux-bias line \cite{Koch2007a,Groszkowski2018} \\ 
		\pyl{t1_inductive}  & Inductive loss due to quasiparticle tunneling in Josephson junction chains that are used to implement superinductances \cite{Pop2014,Smith2020}  \\ 
		\pyl{t1_quasiparticle_tunneling}  & Loss due to quasiparticle tunneling across a single Josephson junction \cite{Pop2014,Smith2020}  \\[2mm]
		\rowcolor{gray!10}
        \scshape $T_{\phi}$ pure-dephasing processes &  \\
        \pyl{tphi_1_over_f_cc} & Dephasing due to $1/f$ critical-current noise (fluctuations of $E_J$ in a Josephson junction) \cite{Ithier2005,Koch2007a,Groszkowski2018} \\ 
		\pyl{tphi_1_over_f_charge}  & Dephasing due to $1/f$ charge noise (fluctuations in charge offset) \cite{Ithier2005,Koch2007a,Groszkowski2018}  \\
		\pyl{tphi_1_over_f_flux}  & Dephasing due to $1/f$ flux noise (fluctuations in the applied magnetic flux) \cite{Ithier2005,Koch2007a,Groszkowski2018}  \\ 
	 \label{table:predefNoiseChannels}
	\end{tabular}
\end{table*}
%\end{widetext}

%\vspace*{3mm}
%{
%%\begin{table*}
    %%\caption{abc}
	%\noindent
	%\small
	%\rowcolors{1}{gray!10}{white}
	%\arrayrulecolor{deepblue} 
	%\setlength{\tabcolsep}{0pt}
	%\setlength\extrarowheight{5pt}
	%\begin{tabular}{p{0.35\columnwidth} >{\raggedright\arraybackslash}p{0.65\columnwidth}}
		%\rowcolor{lightblue}
		%\textbf{noise channel}                       & \textbf{description}                                                   \\\hline
	%\end{tabular}
%%\end{table*}
%}
%\vspace*{3mm}

\subsection{Effective coherence times}
\label{sec:effectiveCoherence}
Coherence times observed in experiments will typically be due to the combined effect of multiple contributing noise channels. \scqubits can easily combine channels and compute effective coherence times or rates. 
In the case of depolarization, the effective coherence time is obtained from
\begin{align}
    \frac{1}{T_{1}^{\rm eff}}  =  \sum_{k} \frac{1}{T_{1}^{k}},
        \label{eq:effDepolarizion}
\end{align}
where the sum runs over all default noise channels, i.e., those methods with names beginning with \pyl{t1}). For each qubit, the included default channels can be listed by calling the qubit's \pyl{effective_noise_channels} method.
To calculate the effective $T_{1}^{\rm eff}$ time based on the default noise channels and their parameters, one simply executes 
\begin{lstlisting}
tmon.t1_effective()
\end{lstlisting}
%Hence to see, for a given qubit, what noise channels will be included in these sorts of effective-coherence time calculations, one can simply execute:
%\begin{lstlisting}
%tune_tmon.t1_effective()
%\end{lstlisting}
Similarly, \scqubits can calculate an effective  dephasing time $T_{2}^{\rm eff}$  (using the method \pyl{t2_effective}). This time scale includes contributions from both pure dephasing as well as depolarization and is defined as
\begin{align}
    \frac{1}{T_{2}^{\rm eff}}  = \sum_{k} \frac{1}{T_{\phi}^{k}} + \frac{1}{2} \sum_{k} \frac{1}{T_{1}^{k}}.
    \label{eq:effDephasing}
\end{align}
Once again the $k$ index cycles over the set of default noise channels.

Both \pyl{t1_effective} as well as \pyl{t2_effective} can be easily customized, so that only select noise channels are incorporated, or calculations be based on specific user-defined parameters. 
For example, a smaller set of channels can be specified by passing a list of channel methods. Further, options shared by all noise channels can be set via the \pyl{common_noise_options} keyword argument, which accepts a dictionary of options.
This is illustrated in the following example, where the temperature is set to $T=0.050\,$K:
\begin{lstlisting}
tmon.t1_effective(
    noise_channels=['t1_charge_impedance',
                    't1_flux_bias_line'],
    common_noise_options=dict(T=0.050)
)
\end{lstlisting}
In addition to \pyl{common_noise_options}, channel-specific noise options can be provided. This is accomplished by replacing the name of the noise-channel method with a tuple of the form \pyl{(channel_name, noise_options)} with \pyl{noise_options} a Python dictionary. In the example below, we calculate an effective $T_{2}^{\rm eff}$, using a non-default value for the $1/f$ flux-noise strength \pyl{A_flux} that internally gets passed to the qubit’s \pyl{tphi_1_over_f_flux} method:
\begin{lstlisting}
tmon.t2_effective(
    noise_channels=['t1_flux_bias_line',
                    't1_capacitive',
                    ('tphi_1_over_f_flux',   
                     dict(A_noise=3e-6))],
    common_noise_options=dict(T=0.050)
)
\end{lstlisting}

\subsection{Coherence visualization}
\label{sec:coherenceVisualization}

A common way to understand and visualize how noise affects a given qubit, is to plot decoherence times as a function of one of the external parameters, such as flux, charge or one of the qubit internal energy parameters, say $E_J$, for example. 
Each qubit provides a flexible method called \pyl{plot_coherence_vs_paramvals}, which facilitates this functionality. 
To provide an overview of the dependence of decoherence properties on, say, flux, the effect of all supported noise channels (as defined by each quibit's \pyl{supported_noise_channels} method), can be visualized in individual plots as follows:
\begin{lstlisting}
tmon.plot_coherence_vs_paramvals(
    param_name='flux', 
    param_vals=np.linspace(-0.5, 0.5, 100));
\end{lstlisting}
Here, \pyl{param_vals} is an array of flux values for which coherence data is generated and plotted. 

The set of plots to be included can be determined by the user and  further customized by either passing various plot options directly to the \pyl{plot_coherence_vs_paramvals} method, or by manipulating the properties of the matplotlib \pyl{Axes} object returned by \pyl{plot_coherence_vs_paramvals}:
\begin{lstlisting}
fig, ax = tmon.plot_coherence_vs_paramvals(
    param_name='flux',
    param_vals=np.linspace(-0.5, 0.5, 100),
    noise_channels=['tphi_1_over_f_flux',     
                    't1_capacitive'],
    scale=1e-3,
    color='red',
    ylabel=r"$\mu$s"
)
ax[0].set_title(r"$t_{\phi}$ from flux noise"); 
ax[0].set_ylim(None, 5e2)
ax[1].set_title(r"$t_1$ from capacitive loss"); 
\end{lstlisting}

\begin{figure*}
	\centering
    \includegraphics[width=0.78\linewidth]{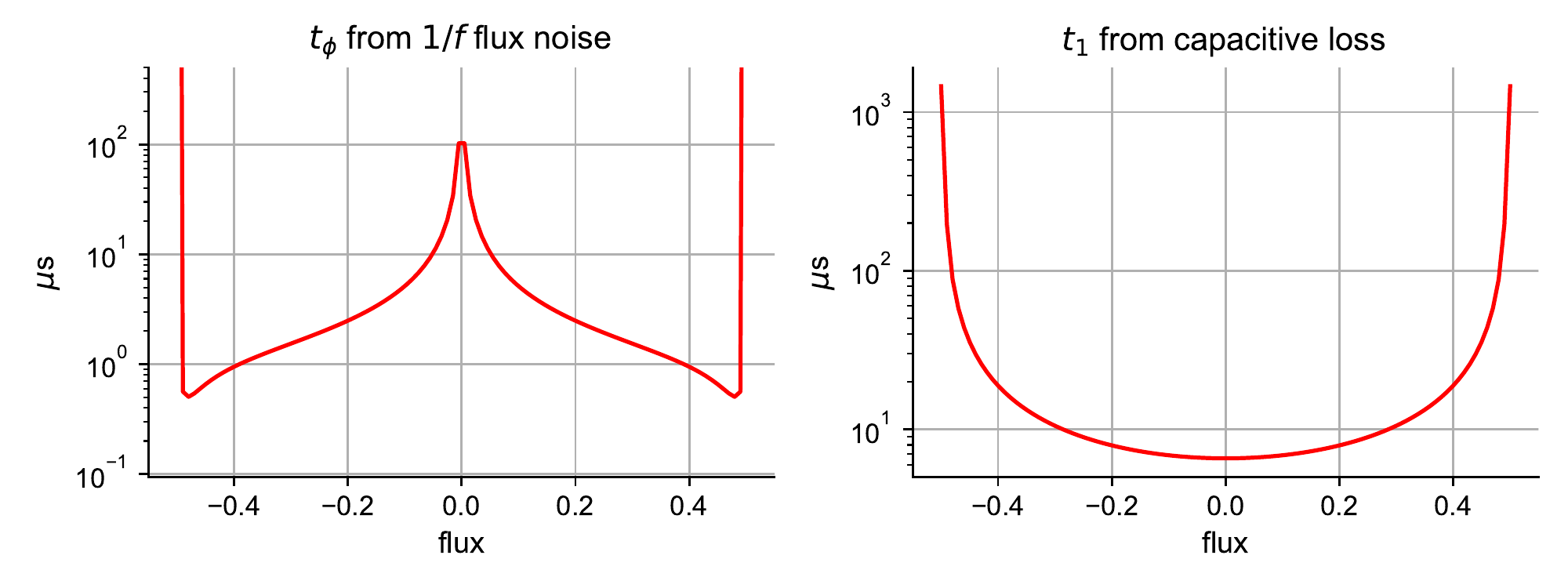}
    \caption{Customized plots showing estimated coherence times of a transmon qubit due to noise from two particular noise channels: \pyl{tphi_1_over_f_flux} and \pyl{t1_capacitive}. Since \scqubits plotting routines  return  standard \pyl{matplotlib} \pyl{Figure} and \pyl{Axes} objects, the plot appearance can be easily modified by the user.    
}
	\label{fig:coherencePlot1}
\end{figure*}
The resulting graphical output (given a \pyl{tmon} object, as defined in Sec.~\ref{sec:overview}), is shown in Fig.~\ref{fig:coherencePlot1}. In this example, only plots for noise channels \pyl{tphi_1_over_f_flux} and \pyl{t1_capacitive} are included. The optional \pyl{scale} argument defines a custom $y$-axis scale factor. Note that the default units of GHz will automatically yield coherence times in units of $n$s. Setting \pyl{scale=1e-3} and adjusting the \pyl{ylabel} of the plot allows us to switch to $\mu$s on the fly. Customizing plots in such ways helps making plots visually appealing and publication-ready, without globally changing  unit setting (see \cref{sec:units}). 

Finally, \scqubits streamlines visualization of the effective noise introduced in \cref{sec:effectiveCoherence} above through the methods \pyl{plot_t1_effective_vs_paramvals} and \pyl{plot_t2_effective_vs_paramvals}. Both methods work in a way analogous to \pyl{plot_coherence_vs_paramvals}, except that now a single plot of the effective coherence time is presented. Provision of an optional \pyl{noise_channels} list now sets the specific noise channels included in the calculation of $T_{1}^{\rm eff}$ and $T_{2}^{\rm eff}$.

\section{Interactive exploration}
\label{sec:interactiveExploration}

Exploring the properties of coupled quantum systems can benefit from visual aides, such as inspection how observables change when system parameters are modified. The \pyl{Explorer} class in \scqubits provides multiple interactive viewgraphs  collecting an important set of information regarding the user-defined system of interest.

The \pyl{Explorer} is based on a \pyl{HilbertSpace} object describing the composite circuit-QED system of interest. As explained in Section 4, \pyl{ParameterSweep} can then be used to record a sweep of an external tuning parameter. This could be, for example, an external magnetic flux, an offset charge, or even a circuit parameter such as a capacitance (difficult to change in-situ in an experiment). 

\subsection{Example: fluxonium coupled to a resonator}
As a concrete example, we consider a system composed of a fluxonium qubit, coupled through its charge operator to the voltage inside a resonator.
The initialization of the composite Hilbert space proceeds as usual: we first define the individual two subsystems that will make up the Hilbert space,
\begin{lstlisting}
fluxonium = scq.Fluxonium(
    EJ=2.55,
    EC=0.72,
    EL=0.12,
    flux=0.0,
    cutoff=110,
    truncated_dim=9
)

osc = scq.Oscillator(E_osc=4.0, truncated_dim=5)
\end{lstlisting}
Here, the \pyl{truncated_dim} parameters are for the hierarchical diagonalization of the composite Hilbert space. For the fluxonium subsystem, \pyl{cutoff} fixes the internal Hilbert space dimension to 110. Once diagonalized, however, only a few eigenstates are usually meant to be retained and included in the composite Hilbert space description. In the example above, the lowest nine states are selected. Similarly, we retain five levels of the resonator, i.e., photon states $n=0,1,\ldots,4$ are included.

\begin{figure*}
	\centering
    \includegraphics[width=0.9\linewidth]{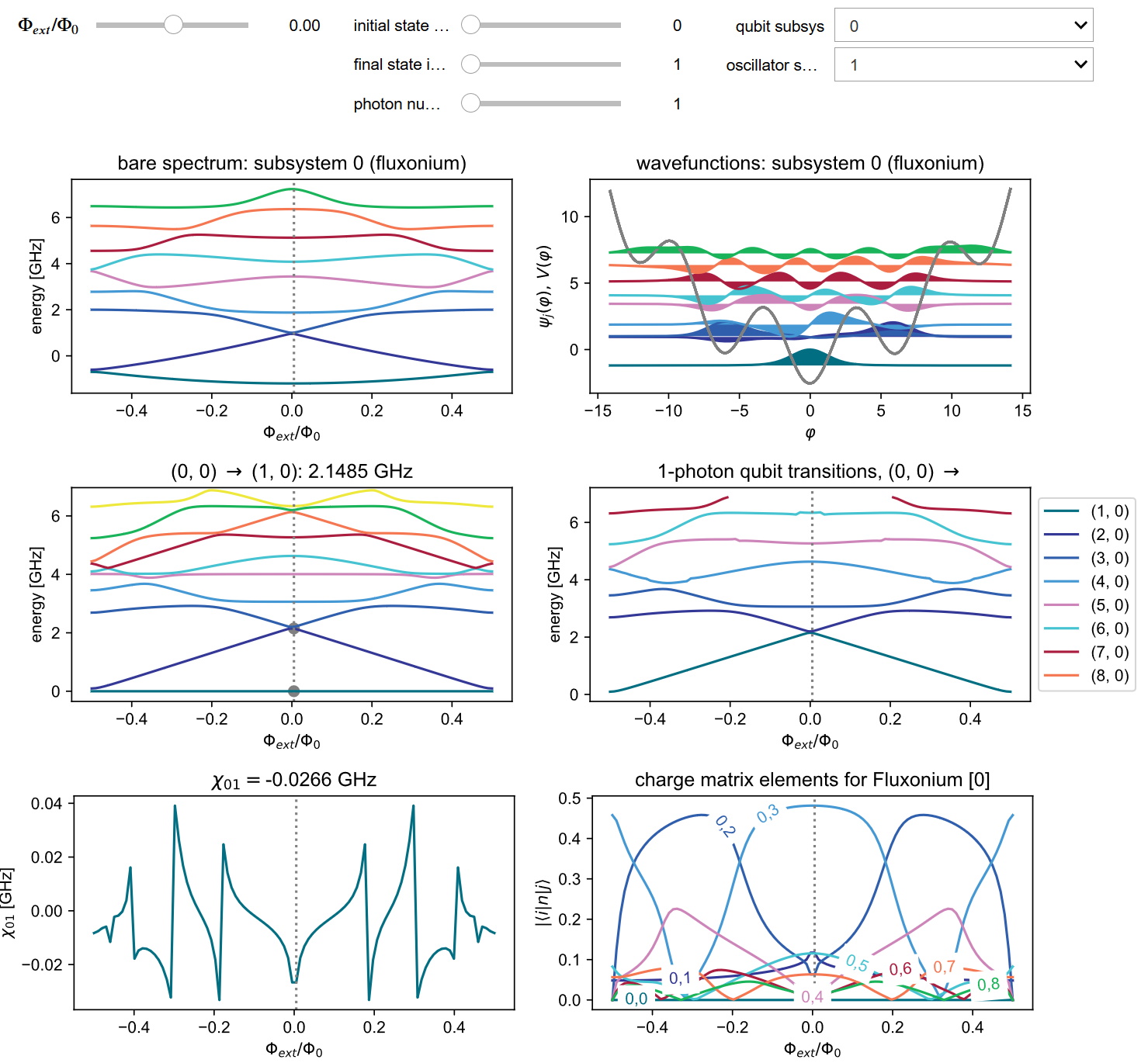}
    \caption{Example output from the interactive \pyl{Explorer} class. The plots shown are (left to right, top to bottom): 1) Bare spectra of the individual qubits,
		2) Wavefunctions of the bare qubits,
		3) Dressed spectrum of the composite Hilbert space,
		4) Spectrum for n-photon qubit transitions, starting from a given initial state,
		5) AC Stark shift $\chi_{01}$ for any of the qubits, and
		6) Charge matrix elements for any of the qubits, using the same initial state as in point 4)}
	\label{fig:fig6}
\end{figure*}

Next, the two subsystems are declared as the two components of a joint Hilbert space:
\begin{lstlisting}
hilbertspace = scq.HilbertSpace([fluxonium, osc])
\end{lstlisting}
The interaction between fluxonium and resonator is of the form $\Hop_\text{int} = g \hat{n} (a+a^\dagger)$, where $\hat{n}$ is the fluxonium's charge operator, \pyl{fluxonium.n_operator}: 
\begin{lstlisting}
hilbertspace.add_interaction(
    g_strength=0.2,
    op1=fluxonium.n_operator,
    op2=osc.creation_operator,
    add_hc=True
)
\end{lstlisting}

As a parameter sweep of common interest, we consider varying the external flux through the fluxonium loop. We create the necessary \pyl{ParameterSweep} object as discussed in Section 4:
\begin{lstlisting}
param_name = r'$\Phi_{ext}/\Phi_0$'
param_vals = np.linspace(-0.5, 0.5, 101)

subsys_update_list = [fluxonium]

def update_hilbertspace(param_val):
    fluxonium.flux = param_val

sweep = scq.ParameterSweep(
    paramvals_by_name={param_name: param_vals},
    evals_count=10,
    hilbertspace=hilbertspace,
    subsys_update_info={param_name: [fluxonium]},
    update_hilbertspace=update_hilbertspace,
)
\end{lstlisting}
At this point, we may start the interactive \pyl{Explorer} class; sample output is shown in Fig.\ \ref{fig:fig6}.

\begin{lstlisting}
explorer = scq.Explorer(
	sweep=sweep,
	evals_count=10
)
explorer.interact()
\end{lstlisting}

\section{Online presence }
\label{sec:onlinePresence}

\scqubits is an open source package, and all of its source code is freely available online under the BSD-3 license. The package is divided into three separate repositories: the first contains the source code of the core package, the second the Sphinx code for the online documentation, and finally the third collects example jupyter notebooks that illustrate how various features of \scqubits can be used.
To install \scqubits, users can either clone the main github repository directly and install via \pyl{pip install .}, or alternatively download and install the package through the PyPI or Anaconda package repositories (see \cite{onlineDocs}). 
For a list of online links to the various github pages containing all the source code, online documentation, live example jupyter notebooks, as well as PyPi and Anaconda package index repository pages, see table \ref{table:links}.
Users, are welcome and encouraged to file bug reports, post comments and suggestions, as well as initiate pull requests on the relevant github pages. Finally, users who find \scqubits useful, are encouraged to cite this paper. The relevant information can be readily accessed by executing the \pyl{cite} function, like so:
\begin{lstlisting}
scq.cite()
\end{lstlisting}

%\begin{table}[t]
\tabcaption{ \scqubits web links to source code, online documentation as well as notebook examples. }
{\footnotesize
	%\rowcolors{1}{gray!10}{white}
	\arrayrulecolor{deepblue} 
	\setlength{\tabcolsep}{0pt}
	\setlength\extrarowheight{5pt}
    \begin{tabular}{p{0.95\columnwidth} }
        %\rowcolor{lightblue}
        %\textbf{class}                       & \textbf{description}                                                   \\\hline
	\rowcolor{gray!10}
    \url{https://github.com/scqubits/scqubits}  -- 
    github repository for \scqubits package source code  \\
    \url{https://github.com/scqubits/scqubits-doc} --
    github repository for \scqubits documentation (Sphinx source code)\\
	\rowcolor{gray!10}
    \url{https://github.com/scqubits/scqubits-examples} -- 
    \scqubits example jupyter notebooks on github \\
    \url{https://scqubits.readthedocs.io/en/latest} --
    \scqubits online documentation  \\
	\rowcolor{gray!10}
    \url{https://mybinder.org/v2/gh/scqubits/scqubits-examples/released} -- 
    \scqubits example jupyter notebooks live demo \\
    \url{https://pypi.org/project/qutip/} --
    \scqubits PyPI package repository \scqubits page \\
	\rowcolor{gray!10}
    \url{https://anaconda.org/conda-forge/scqubits} --
    \scqubits Anaconda package repository \scqubits page 
    \label{table:links}
    \end{tabular}
}
%\end{table}

\section{Conclusions}
\label{sec:conclusions}
With this paper, we have introduced the \scqubits library: an open-source Python toolbox enabling the simulation of superconducting qubits -- both at the single-qubit level, and at the level of composite quantum systems consisting of multiple qubits and resonators. The current functionality of the library encompasses a broad range of computational tasks and visualization tools commonly used in research involving superconducting qubits. While maintaining the existing interface, future work will aim to extend the scope of the library, for example by including a systematic workflow for fitting experimental two-tone spectroscopy data, analyzing custom circuits defined by the user, implementing newer qubit designs, as well adding to the list of predefined noise channels in order to extend coherence time estimations.

\section*{Acknowledgments}
We thank P.\ Aumann, E.\ Blackwell, S.\ Chakram, F.\ Hassani, Z.\ Huang, N.\ Irons, P.\ Mundada, D. Schuster, J.\ Sung, S.\ Wang, D.\ Weiss, X.\ You, and A.\ Zheng for code contributions and bug reports. Continuing development of \scqubits is  currently supported by the AFOSR under grant FA9550-20-1-0271. Initial work on the package was in part supported by the ARO under grants W911NF-15-1-0421 and W911NF-19-1-0016, and by the Northwestern-Fermilab Center for Applied Physics and Superconducting Technologies (CAPST).

\bibliographystyle{plainnat}
\bibliography{library}

%%%%%%%%%%%%%%%%%%%%% APPENDIX %%%%%%%%%%%%%%%%%%%%%%%%%%%%%%%%
\clearpage
%\onecolumngrid
\appendix
\section*{Appendix}
\renewcommand{\thesubsection}{A.\Roman{subsection}}
\renewcommand{\thesection}{}
\setcounter{equation}{0}
\numberwithin{equation}{section}
\renewcommand{\theequation}{A.\arabic{equation}}
\setcounter{figure}{0}
\renewcommand{\thefigure}{A.\arabic{figure}}
\renewcommand{\theHfigure}{A.\arabic{figure}}

%%%%%%%%%%%%%%%%%%%%%%%%%%%%%%%%%%%%%%%%%%%%%%%%%%%%%%%%%%%%%%%%%%%%%%%%%%%%%%%%%%%%%%%%%%%%%%%%%%%%

\section{Superconducting qubit and oscillator classes}
\label{app:qubitClasses}
\scqubits currently implements several common types of superconducting qubits along with a linear and non-linear oscillators, as well as a generic qubit (i.e., a simple two-level system), each realized as a Python class\footnote{More qubit types will be added in the future. See online documentation for the latest information \cite{onlineDocs}.}. 
A brief summary is shown in the following table \cite{onlineDocs}:
%\vspace{3mm}
{
	%\noindent
	\small
	\rowcolors{1}{gray!10}{white}
	\arrayrulecolor{deepblue} 
	\setlength{\tabcolsep}{0pt}
	\setlength\extrarowheight{5pt}
	\begin{tabular}{p{0.35\columnwidth} >{\raggedright\arraybackslash}p{0.65\columnwidth}}
		\rowcolor{lightblue}
		\textbf{class}                       & \textbf{description}                                                   \\\hline
		\pyl{Transmon}, \pyl{TunableTransmon} & Transmon qubit or \mbox{Cooper pair box} \cite{Nakamura1999,Koch2007a} \\
		\pyl{Fluxonium}                      & fluxonium qubit \cite{Manucharyan2009}                                 \\
		\pyl{FluxQubit}                      & 3-junction flux qubit \cite{Mooij1999}                                 \\
		\pyl{ZeroPi}                         & 0-$\pi$ qubit (symmetric) \cite{Brooks2013,Dempster2014a}              \\
		\pyl{FullZeroPi}                     & 0-$\pi$ qubit (coupled to $\zeta$-mode) \cite{Dempster2014a}            \\
        \pyl{Cos2PhiQubit}                   & $\cos2\phi$ qubit \cite{Smith2020} \\
        \pyl{Oscillator}                   & Quantum harmonic oscillator \\
        \pyl{KerrOscillator}                   & Nonlinear Kerr oscillator  \\
        \pyl{GenericQubit}                   & A two-level system 
	\end{tabular}
}
\vspace*{3mm}

All the qubit classes (except for \pyl{GenericQubit}) define a number of important methods that can be used for diagonalization, computation of matrix elements and spectral data, as well as for plotting. A few of these are summarized in Table \ref{table:selctedQubitMethods}. 

Besides these methods, each superconducting qubit class also implements a predefined number of quantum operators which can simplify doing various calculations. These may include the phase $\phiop$ or number $\nop$ operators, but also more specialized ones such as $\cos(\phiop)$ and the like. See the API documentation \cite{APIdocs}, for a comprehensive list.

In the following sections we describe the qubit and oscillator classes that \scqubits implements in more detail: present their circuit diagrams (where applicable), give definitions of the respective Hamiltonians, and briefly discuss their numerical implementation in the library.

\subsection{Harmonic and Kerr oscillators }
\label{app:oscillators}

The most basic quantum systems that can be realized in \scqubits are harmonic and Kerr resonators with Hamiltonians
\begin{align}
    \Hop_\text{osc}=E_\text{osc} \aop \dg \aop ,
\end{align}
and
\begin{align}
    \Hop_\text{Kerr}=E_\text{osc} \aop \dg \aop - K \aop \dg \aop \dg  \aop \aop,
\end{align}
respectively, where $\aop$ corresponds to a standard bosonic lowering operator, while $E_\text{osc}$ and $K$ are the oscillator and Kerr energies, respectively. 
Example initialization code for both cases is shown below. For a harmonic oscillator, we have
\begin{lstlisting}
osc=scq.Oscillator(E_osc=5)
\end{lstlisting}
while for the Kerr oscillator,
\begin{lstlisting}
kerr=scq.KerrOscillator(E_osc=5, K=0.1, 
                        l_osc=0.1)
\end{lstlisting}
Note that the oscillator length  \pyl{l_osc} is an optional parameter in both oscillator classes, which if not given, is set to \pyl{None}. It, however, has to be provided by the user for the oscillator classes to define phase $\phiop = l_\text{osc} (\aop^{\dagger} + \aop)/\sqrt{2}$ and number $\noper = i  (\aop^{\dagger} - \aop ) / \sqrt{2} l_\text{osc} $ operators\footnote{For a quantum harmonic oscillator represented by a Hamiltonian $H=E_{\text{kin}} \Pop^{2} + E_{\text{pot}} \Xop^{2} $, the oscillator length is defined as $l_{\text{osc}}=(E_{\text{kin}} / E_{\text{pot}})^{1/4}$.}, which are implemented in methods \pyl{phi_operators} as well as \pyl{n_operator} respectively.

\subsection{Generic Qubit}
\label{app:genericQubit}

\scqubits implements a generic qubit class called \pyl{GenericQubit}, that corresponds to a simple two-level system with a Hamiltonian
\begin{align}
    \Hop &=  \frac{E}{2} \hat{\sigma}_{z},
\end{align}
where $E$ is the qubit's energy.
This class implements a set of Pauli operators $\hat{\sigma}_{k}$, with $k=\{x,y,z\}$, as well as lowering and raising operators $\hat{\sigma}_{\pm}$, with methods names such as \pyl{sx_operator}, \pyl{sm_operator}, etc.. 
To initialize a \pyl{GenericQubit} object, one provides its energy:
\begin{lstlisting}
qubit = scq.GenericQubit(E=5.5)
\end{lstlisting}

\subsection{Transmon qubit}
 \label{app:transmon}
The Cooper pair box \cite{Nakamura1999} and transmon qubit \cite{Koch2007a} share the same underlying circuit composed of a single Josephson junction and a parallel capacitance which may either be the pure junction capacitance, or include an external shunt capacitance:
\begin{center}
	\includegraphics[width=0.45\columnwidth]{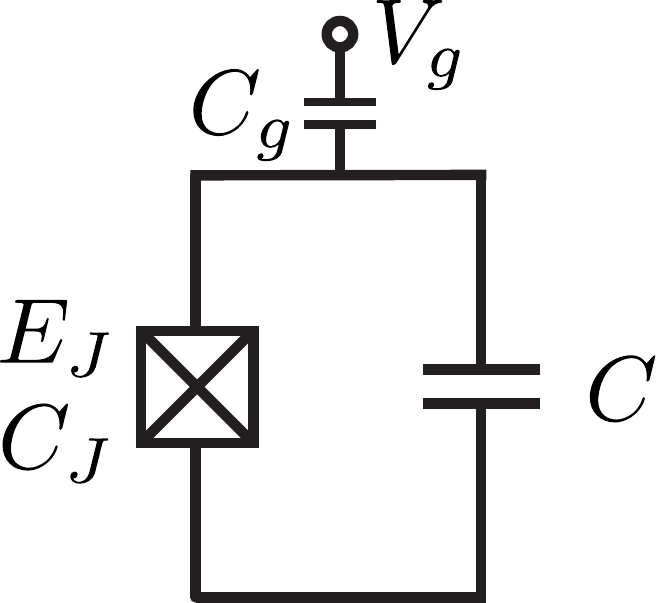}
\end{center}
 The circuit Hamiltonian can be written as
\begin{align}
	%H & = 4E_C(-i\partial_\phi - n_g)^2 - E_J \cos\phiop \\\nonumber
    \Hop_{\text{transmon}}  & =4E_C(\noper-n_g)^2 - E_J \cos\phiop,
\end{align}
where $\nop$ ($\phiop$) is the charge (phase) operator\footnote{Since the Hamiltonian is periodic in $\phiop$, we stress that only periodic functions of $\phiop$ are formally defined.}
%$\left[\phiop, \nop \right] \ne i$. \Jens{That misses the point... The problem is not at all the commutator. The problem is that $\phiop$ itself does not exist.}}.
%\begin{align}
	%H & = 4E_C(-i\partial_\phi - n_g)^2 - E_J \cos\phiop \\\nonumber
	  %& =4E_\text{C}(\hat{n}-n_g)^2+\frac{1}{2}E_\text{J}\sum_{n=-\infty}^\infty (|n\rangle\langle n+1|+\text{h.c.}), 
%\end{align}
Here $E_C=e^2/2C_\Sigma$ is the charging energy associated with the combined capacitances of the junction, the shunt capacitor, and any additional ground capacitance and/or capacitance to a charge bias line,  $C_\Sigma = C_J + C + C_g$. The Josephson energy of the junction is related to its critical current via $E_J=I_c\Phi_0/2\pi$. The quantity $n_g$ is the dimensionless offset charge capturing the capacitive coupling to a bias voltage source as well as electric-potential fluctuations of the environment. 

Internally, the \pyl{Transmon} class employs the charge-basis representation to construct the Hamiltonian matrix. This matrix is infinite-dimensional, in principle, and must hence be truncated. To this end, a charge-number cutoff \pyl{ncut} is introduced. Given this cutoff, the included charge states $|n\rangle$ are within the range -\pyl{ncut}$\le n \le$\pyl{ncut}. The cutoff must be chosen sufficiently large to avoid truncation errors\footnote{For low-lying transmon wavefunctions ($E_J/E_C\gg1$), there is a simple cutoff criterion: The ground state is close to Gaussian with standard deviation $\sigma = (8E_C/E_J)^{1/4}$. Treating $n$ as continuous and assuming an $n_g$ of order 1, Fourier transform of the ground state yields $\psi_0(n)$ which is also close to a Gaussian, here with standard deviation $\sigma' = (E_J/8E_C)^{1/4}$. Using a $3\sigma$ estimate, one finds that \pyl{ncut} should be no smaller than
$
\texttt{ncut}_\text{min} \approx \left\lceil 2(E_J/E_C)^{1/4} \right\rceil.$}.

%\begin{widetext}
\begin{table*}[t]
	\tabcaption{A summary of a few selected methods shared by all the qubit classes. For full information including the call signatures of these methods, see the API documentation \cite{APIdocs}. \label{table:qub1}}
	\small
	%\rowcolors{1}{gray!10}{white}
	\arrayrulecolor{deepblue} 
	\setlength{\tabcolsep}{0pt}
	\setlength\extrarowheight{5pt}
	\begin{tabular}{p{0.33\textwidth} >{\raggedright\arraybackslash}p{0.67\textwidth}}
		\rowcolor{lightblue}
		\textbf{Qubit class method}             & \textbf{description}                                                                                 \\\hline
		\rowcolor{gray!10}
		\scshape Basics                         &                                                                                                      \\
		%\pyl{hiltertdim()} & dimension of circuit Hilbert space\\
		\pyl{hamiltonian}                     & Hamiltonian matrix (in basis specific to each qubit)                                                 \\
		\pyl{eigenvals}                       & Eigenvalues of the qubit Hamiltonian                                                                 \\
		\pyl{eigensys}                        & Eigenvalues and eigenvectors of the qubit Hamiltonian                                                \\
		\rowcolor{gray!10}
		\scshape Matrix elements, Spectral data &                                                                                                      \\
		\pyl{wavefunction}                  & Wavefunction of qubit (in basis dependent on each qubit)                                             \\ 
		\pyl{matrixelement_table}           & Matrix elements for a specified qubit operator, with respect to a set of qubit eigenstates           \\
		\pyl{get_spectrum_vs_paramvals}     & Compute eigenenergies and eigenstates as a function of a specified external parameter                \\
		\pyl{get_matelements_vs_paramvals}  & Compute matrix elements for specified qubit operator as a function of a specified external parameter \\
		\rowcolor{gray!10}
		\scshape{Plotting methods}              &                                                                                                      \\
		\pyl{plot_wavefunction}             & Plot wavefunction(s) of qubit                                                                        \\
		\pyl{plot_evals_vs_paramvals}       & Plot of energy eigenvalues as a function of specified external parameter                             \\
		\pyl{plot_matrixelements}           & Combined 3d bar plot and 2d plot of matrix elements for specified qubit operator                     \\
		\pyl{plot_matelem_vs_paramvals}     & Plot of matrix elements for specified qubit operator as a function of external parameter             
        \label{table:selctedQubitMethods}
	\end{tabular}
\end{table*}[t]

An example initialization code of a Transmon qubit is shown below
\begin{lstlisting}
transmon = scq.Transmon(EJ=30.02, EC=0.2, 
                        ng=0.0, ncut=101)
\end{lstlisting}

\subsection{Fluxonium qubit}
 \label{app:fluxonium}

The circuit of the fluxonium qubit \cite{Manucharyan2009} consists of a Josephson junction shunted by a large inductor:
\begin{center}
	\includegraphics[width=0.45\columnwidth]{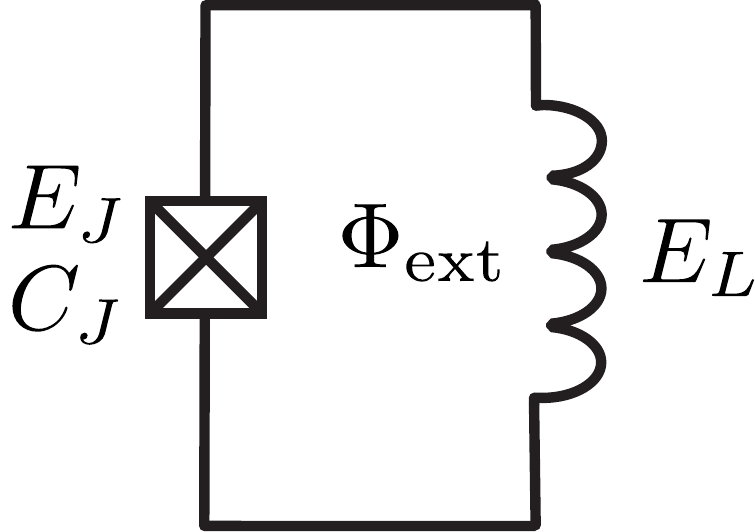}
\end{center}
The resulting qubit Hamiltonian takes the form 
%\begin{align}
	%H & =-4E_\text{C}\partial_\varphiop^2-E_\text{J}\cos(\varphiop-\varphi_\text{ext}) +\frac{1}{2}E_L\varphiop^2, 
%\end{align}
\begin{align}
    \Hop_{\text{fluxonium}} & =4E_\text{C}\nop^2-E_\text{J}\cos(\varphiop-\varphi_\text{ext}) +\frac{1}{2}E_L\varphiop^2, 
\end{align}
where $E_C=e^2/2C_{J}$ and $E_J$ are the charging and Josephson energies of the junction, and $E_L=(\frac{\Phi_0}{2\pi})^2/L$ is the inductive energy.
The Hilbert-space basis used by \scqubits for construction of the Hamiltonian is the harmonic-oscillator basis associated with the inductor and junction capacitor. The Hamiltonian can be rewritten in this basis by employing the usual ladder operators $a$, $a^\dag$,
\begin{align}\label{Hflux}
    \Hop_\text{fluxonium} & = \sqrt{8E_L E_C}\, a^\dag a                                                                                                \\\nonumber
	  & \quad -\frac{E_J}{2} e^{-i\varphi_\text{ext}} \exp\bigg[\frac{i\varphi_\text{osc}}{\sqrt{2}}(a+ a^\dag)\bigg] + \text{h.c.} 
\end{align}
Here, we have rewritten $\cos(\varphiop-\varphi_\text{ext})= \frac{1}{2} e^{-i\varphi_\text{ext}}e^{i\varphiop} + \text{h.c.}$, and used $\varphiop=\frac{\varphi_\text{osc}}{\sqrt{2}}(\aop+\aop \dg)$ with $\varphi_\text{osc}=(8E_C/E_L)^{1/4}$ denoting the oscillator length. Numerical evaluation of the matrix exponential in \eqref{Hflux} is carried out via \pyl{scipy.linalg.expm()}.

%\begin{table}
%\caption{Essential attributes and operator methods for the \pyl{Fluxonium} qubit.}
    %\small
    %\rowcolors{1}{gray!10}{white}
    %\arrayrulecolor{deepblue} 
    %\setlength{\tabcolsep}{0pt}
    %\setlength\extrarowheight{5pt}
%\begin{tabular}{p{0.35\columnwidth} >{\raggedright\arraybackslash}p{0.65\columnwidth}}
          %\rowcolor{lightblue}
	%\textbf{\pyl{Fluxonium} \mbox{attributes}} & \textbf{description}\\\hline
	%\pyl{EJ} & Josephson energy $E_J$\\
	%\pyl{EC} & charging energy $E_C=e^2/2C$\\
	%\pyl{EL} & inductive energy $E_L=(\frac{\Phi_0}{2\pi})^2/L$\\
	%\pyl{flux} & external flux $\Phi_\text{ext}/\Phi_0$\\
	%\pyl{cutoff} & $LC$ oscillator cutoff
%\end{tabular}
%%
%\mbox{}\\[3mm]
%%
%\begin{tabular}{p{0.45\columnwidth} >{\raggedright\arraybackslash}p{0.55\columnwidth}}\rowcolor{lightblue}
	%\textbf{\pyl{Fluxonium} operators} & \textbf{description}\\\hline
	%\pyl{cos_phi_operator()} &  $\cos\varphi$ in $LC$ osc.\ basis\\
	%\pyl{sin_phi_operator()} & $\sin\varphi$ in $LC$ osc.\ basis\\
	%\pyl{exp_i_phi_operator()} & $e^{i\varphi}$ in $LC$ osc.\ basis\\
	%\pyl{n_operator()} & $n=-i \partial_\varphi$ in $LC$ osc.\ basis\\
	%\pyl{phi_operator()} & $\varphi$ in $LC$ oscillator basis
%\end{tabular}
%\end{table}

It must be noted that the harmonic-oscillator basis is not well-adapted to fluxonium eigenstates that localize in individual wells of the cosine contribution to the potential. Consequently, the inevitable truncation in the harmonic-oscillator basis must generally proceed with caution, and the cutoff number \pyl{cutoff} be chosen sufficiently large for convergence.

An example initialization code of a Fluxonium qubit is show below
\begin{lstlisting}
fluxonium = scq.Fluxonium(EJ=5.7, EC=1.2, 
                          EL=1.0, cutoff = 150, 
                          flux = 0.5,
                          truncated_dim=10)
\end{lstlisting}

%At the moment we're not listing all ops for all qubits
%The operators defined in the \pyl{Fluxonium} class are:

\subsection{Flux qubit}
 \label{app:fluxQubit}

 The 3-junction flux qubit that \scqubits implements, was first proposed in \cite{orlando1999superconducting}. Its circuit consists of 3 Josephson junctions in a loop that is threaded with an external flux $\Phi_{\text{ext}}$.
\begin{center}
	\includegraphics[width=0.7\columnwidth]{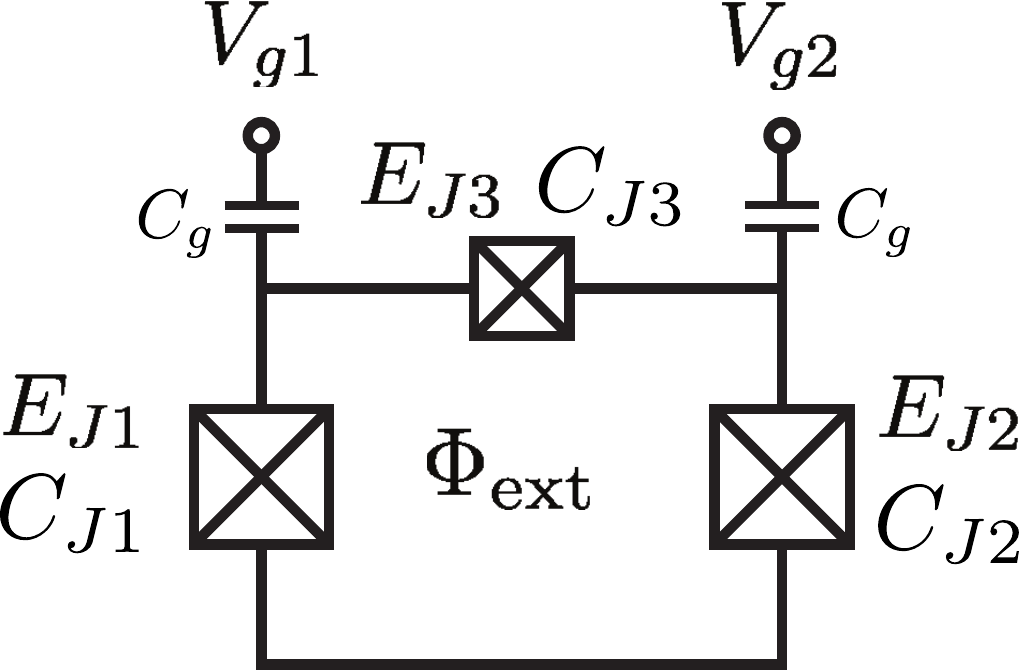}
\end{center}
$E_{Jk}$ ($C_{Jk}$) with $k \in \{1,2,3\}$ are the Josephson energies (capacitances) associated with each junction.
For normal qubit operations one junction is chosen to be smaller than the other two. 
The effective Hamiltonian of such a flux qubit is described by 
\begin{align}
    \Hop_{\text{flux}} &= \sum_{j,k=1}^2 4(\mathsf{E}_\text{C})_{jk}(\noper_{j}-n_{gj})(\noper_{j}-n_{gk})  \nonumber  \\
	  &  -\sum_{k=1}^2 E_{Jk}\cos\varphiop_{k}                                                                           
      - E_{J3}\cos(\varphiop_1 - \varphiop_2 + \varphi_\text{ext}). 
    %H_{\text{flux}} &= \sum_{j,k=1}^2 4(\mathsf{E}_\text{C})_{jk}(- i \partial_{\varphi_j}-n_{gj})(- i \partial_{\varphi_k} -n_{gk})  \nonumber  \\
	  %& \qquad -\sum_{k=1}^2 E_{Jk}\cos\varphiop_{k}                                                                           
      %- E_{J3}\cos(\varphiop_1 - \varphiop_2 + \varphi_\text{ext}). 
	  %& =\sum_{j,k=1}^2 4(\mathsf{E}_\text{C})_{ij}(\hat{n}_j-n_{gj})(\hat{n}_k -n_{gk})                                     \\\nonumber
	  %& \qquad -\sum_{k=1}^2 \frac{E_{Jk}}{2} \left( | n_k\rangle \langle n_k+1| +\text{h.c.} \right)                        \\\nonumber
	  %& \qquad -\frac{E_{J3}}{2} \left(e^{i \varphi_\text{ext}} |n_1,n_2\rangle \langle n_1 -1, n_2+1| + \text{h.c.} \right) 
      \label{eq:fluxQubitH}
\end{align}
In the above expression, $\mathsf{E}_\text{C}$ represents a charging energy matrix (which includes effects of small capacitors $C_{g}$), while $n_{gj}$ (with $j \in \{1,2\}$) are the charge offsets. The two degrees of freedom are represented by $\varphiop_{1}$ and $\varphiop_{2}$, with their conjugates charge operators $\noper_{1}$ and $\noper_{2}$ respectively.  
Numerical diagonalization is performed in the charge basis for both $\varphiop_{1}$ and $\varphiop_{2}$ (see discussion in \ref{app:transmon}).

A sample initialization code of the Flux qubit, where the third junction is assumed to be smaller than the other two by a factor of $\alpha$, is shown below
\begin{lstlisting}
EJ = 35.0
alpha = 0.6
fluxqubit = scq.FluxQubit(
    EJ1 = EJ,
    EJ2 = EJ,
    EJ3 = alpha*EJ,
    ECJ1 = 1.0,
    ECJ2 = 1.0,
    ECJ3 = 1.0/alpha,
    ECg1 = 50.0,
    ECg2 = 50.0,
    ng1 = 0.0,
    ng2 = 0.0,
    flux = 0.5,
    ncut = 10
)
\end{lstlisting}

\subsection{0-$\pi$ qubit}
 \label{app:zeroPi}

The $0-\pi$ qubit was first proposed in Ref. \cite{Brooks2013}. Its circuit consists of two Josephson junctions, two capacitors as well as two superinductors arranged in the following form:
\begin{center}
	\includegraphics[width=0.60\columnwidth]{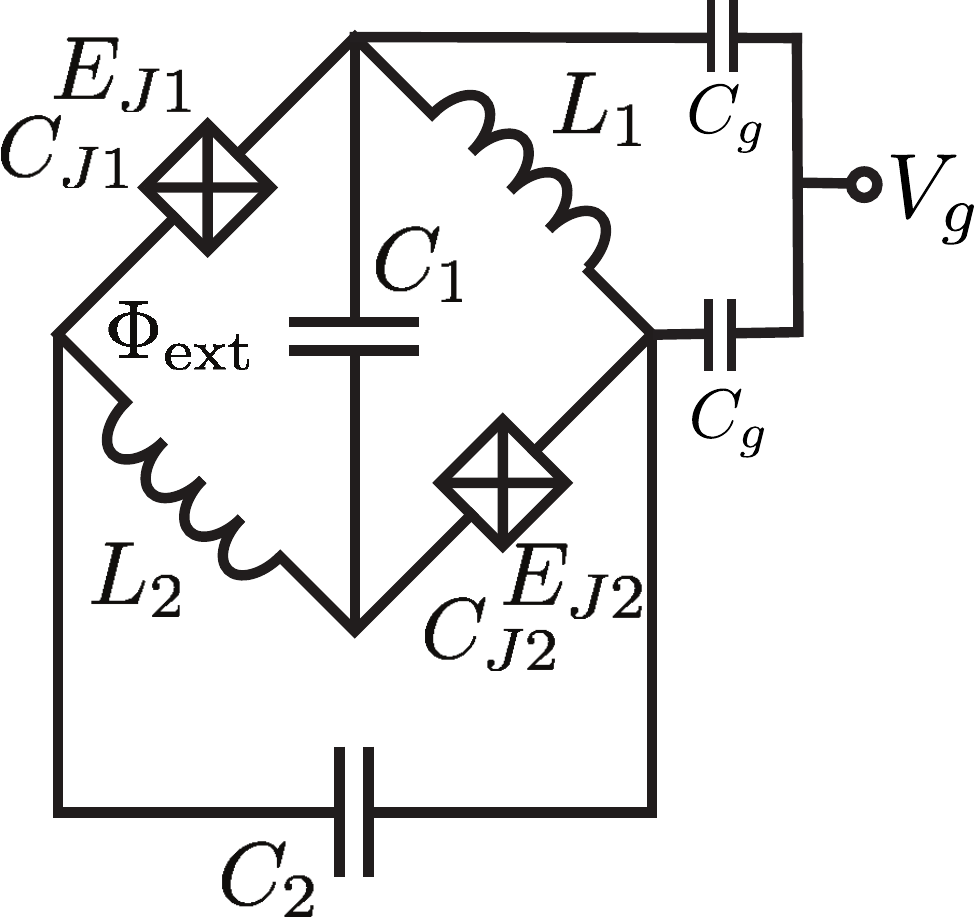}
\end{center}
The behavior of this qubit has been shown to strongly depend on the presence of parameter disorder \cite{Dempster2014a,Groszkowski2018}. In particular, when the two (non-junction) capacitive or inductive energies are not identical (i.e., $C_{1} \ne C_{2}$ or $L_{1} \ne L_{2}$), the core qubit degrees of freedom $\thetaop$ and $\phiop$ end up coupling to a low-frequency harmonic mode $\zetaop$, which would stay uncoupled when there is no disorder in in these parameters. 
For this reason, \scqubits implements two separate classes that can be used for modeling of $0-\pi$ qubits. 
The first, \pyl{ZeroPi}, assumes both of the inductors as well as the (non-junction) capacitors are identical and hence only includes the $\thetaop$ and $\phiop$ degrees of freedom. 
The second, \pyl{FullZeroPi}, allows for small parameter disorder in the superinductnaces $L_{k}$ as well as the (non-junction) capacitors $C_{k}$, resulting in a three degrees of freedom system, which now also includes the harmonic $\zetaop$ mode. 
Note that the \pyl{ZeroPi} class does allow disorder in the Josephson junctions (i.e., $E_{J1}$ and $C_{J1}$ do not have to be identical to  $E_{J2}$ and $C_{J2}$), as this kind of disorder still leaves the $\zetaop$ mode decoupled from $\thetaop$ and $\phiop$.  

To quantify disorder in any parameter $X$ (with $X \in \{C, E_{L}, C_{J}, E_{J} \}$, we define
\begin{align}
    X &= \frac{X_{1} + X_{2}}{2} \quad \quad dX = \frac{X_{1} - X_{2}}{X}.
\end{align}
Then, in the limit of small disorder $dX$ \cite{Dempster2014a,Groszkowski2018}, a general $0-\pi$ Hamiltonian can be approximated by 
\begin{align}
    \Hop_\text{tot} &=  \Hop_{0-\pi} + \Hop_\text{int} + \Hop_\zeta
\end{align}
with 
\begin{align}
\label{eq:zeropiH0}
\Hop_{0-\pi} =&  2E_\text{CJ}\noper_{\phi}^2+2E_{\text{C}\Sigma}(\noper_{\theta}+n_g)^2  \\\nonumber
&- 2 E_{J}\cos\thetaop \cos \left( \phi - \frac{ \varphi_\text{ext}}{2} \right) + E_{L} \phiop^{2} \\\nonumber
&- 2E_{C\Sigma}dC_J\, \noper_\phi \noper_\theta \\\nonumber 
&+  E_{J} dE_{J} \sin\thetaop\sin \left(\phiop-\frac{\phi_\text{ext}}{2}\right),   
\end{align}
along with 
\begin{align}
\Hop_\text{int} =& -2E_{C\Sigma}dC\,\noper_\theta\noper_\zeta + E_L dE_L \phiop\,\zetaop,
%\Hop_\text{int} =& 2E_{C\Sigma}dC\,\partial_\theta\partial_\zeta + E_L dE_L \phiop\,\zetaop,
\label{eq:zeropiHint}
\end{align}
and
\begin{align}
    \Hop_\zeta =& \omega_\zeta \aop \dg \aop .
\label{eq:zeropiHzeta}
\end{align}
The quantity $n_{g}$ is the charge offset of $\noper_\theta$, while $\varphi_\text{ext}=2\pi \Phi_\text{ext}/\Phi_{0}$. 
The class \pyl{ZeroPi} only implements $\Hop_{0-\pi}$, as it clear from the above description, that when $dE_{L}=dC=0$, $\Hop_{\text{int}}=0$, 
%thus making the $\zetaop$ mode uncoupled from $\Hop_{0-\pi}$
, while \pyl{FullZeroPi} includes all three terms of $\Hop_\text{tot}$. 
Internally, we use charge basis to describe the $\thetaop$ degree of freedom (see \ref{app:transmon}), phase basis for the $\phiop$ degree of freedom, and finally $\zetaop$ is modeled using harmonic basis (see \ref{app:fluxonium}). 
For problems where the disorder $dC$ and $dE_{L}$ can be neglected, it is strongly recommended that \pyl{ZeroPi} is used, as the performance penalty from including the physics of the $\zetaop$ mode can be substantial.

Below is sample code showing an initialization of a $0-\pi$ qubit. Here, we include a $5\%$ disorder in the Josephson junction energies, but assume there is no disorder in the superinductors and (non-junction) capacitors, allowing us to use the \pyl{ZeroPi} class:
\begin{lstlisting} % using parameter set 2 from Groszkowski et al.
phi_grid = scq.Grid1d(-8*np.pi, 8*np.pi, 200)
zeropi_dis = scq.ZeroPi(
    grid = phi_grid,
    EJ   = 10.0,
    dEJ = 0.05,
    EL   = 0.04,
    ECJ  = 20.0,
    dCJ = 0.05,
    EC = 0.04,
    ng   = 0.3,
    flux = 0.2,
    ncut = 30
)
\end{lstlisting}

\subsection{$\cos2\phi$ qubit}
 \label{app:cos2phi}

The $\cos2\phi$ qubit was proposed in \cite{Smith2020}. 
Its circuit includes a superconducting loop, threaded by an external flux $\Phi_{\text{ext}}$, which consists of two superinductors and two Josephson junctions, with an appropriately placed shunt capacitance:
\begin{center}
    \includegraphics[width=0.75\columnwidth]{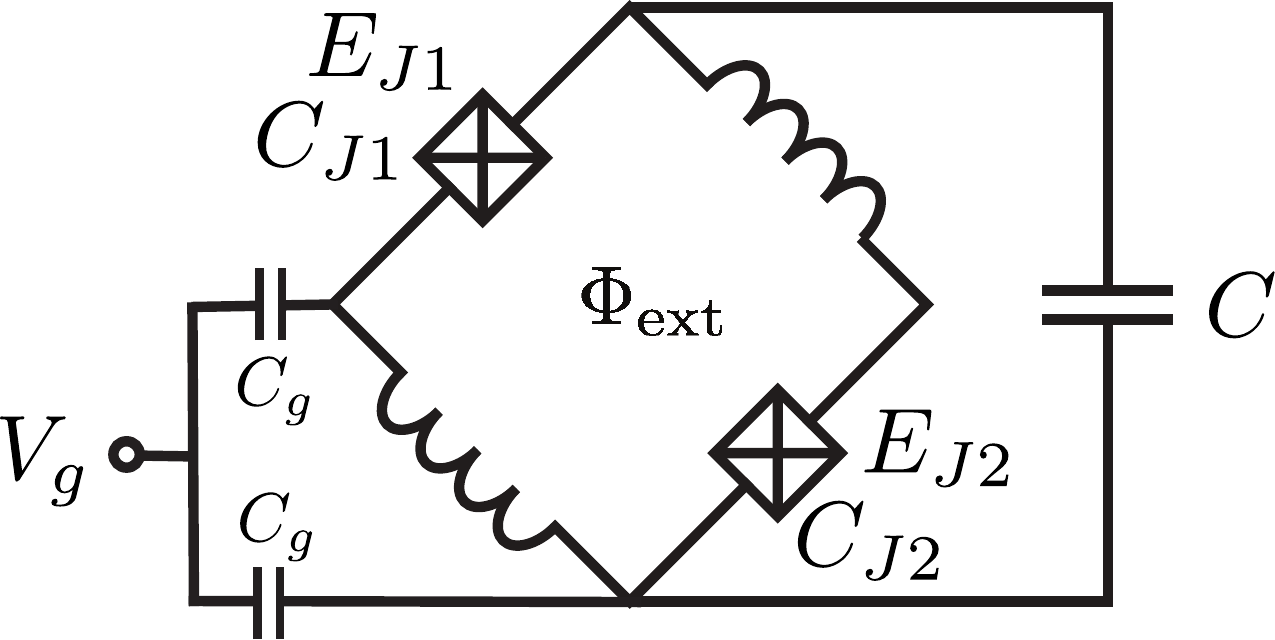}
\end{center}
Such topology can be engineered to only allow pairs of cooper pairs to tunnel, leading a level of intrinsic protection from noise \cite{Smith2020}. 
The Hamiltonian of a $\cos2\phi$ qubit can be  written as 
%\begin{align}
    %\Hop_{\cos2-\phi} = & \,2 E_\text{CJ}'\noper_\phi^2 + 2 E_\text{CJ}' (\noper_\theta - n_\text{g} - \noper_\zeta)^2 + 4 E_\text{C} \noper_\zeta^2 \nonumber \\
    %& + E_\text{L}'(\phiop - \pi\Phi_\text{ext}/\Phi_0)^2 \\\nonumber 
    %&+ E_\text{L}' \zetaop^2 - 2 E_\text{J}\cos{\thetaop}\cos{\phiop} \nonumber \\
    %& + 2 dE_\text{J} E_\text{J}\sin{\thetaop}\sin{\phiop} \nonumber \\
    %& - 4 dC_\text{J} E_\text{CJ}' \noper_\phi (\noper_\theta - n_\text{g}-n_\zeta) \nonumber \\
    %& + dL E_\text{L}'(2\phiop - \varphi_\text{ext})\zeta ,
%\end{align}
\begin{align}
    \Hop_{\cos2\phi} = & \,2 E_{CJ}'\noper_\phi^2 + 2 E_{CJ}' (\noper_\theta - n_{g} - \noper_\zeta)^2 + 4 E_{C} \noper_\zeta^2 \nonumber \\
    & + E_{L}'(\phiop - \pi\Phi_\text{ext}/\Phi_0)^2 \\\nonumber 
    &+ E_{L}' \zetaop^2 - 2 E_{J}\cos{\thetaop}\cos{\phiop} \nonumber \\
    & + 2 dE_{J} E_{J}\sin{\thetaop}\sin{\phiop} \nonumber \\ \nonumber
    & - 4 dC_{J} E_{CJ}' \noper_\phi (\noper_\theta - n_{g}-n_\zeta) \nonumber \\ \nonumber
    & + dL E_{L}'(2\phiop - \varphi_\text{ext})\zeta,
\end{align}
with $E_{CJ}' = E_{CJ} / (1 - dC_{J})^2$ and $E_{L}' = E_{L} / (1 - dL)^2$. 
Parameters of the form $dX$ represent disorder\footnote{Note the difference in the definition of disorder here versus the definition for the $0-\pi$ qubit. 
%In particular, in the case of $\cos2\phi$, we have $dX = (X_{1} - X_{2})/2X$.
The definition of parameter disorder in the $\cos2\phi$ qubit used by \scqubits follows the notation in  \cite{Smith2020}.}. 
In particular, we have 
\begin{align}
    dX = \frac{X_{1} - X_{2}}{X_{1} + X_{2}}, 
\end{align}
for $X \in \{L, C_{J}\}$, with $L$ ($C_{J}$) being the superinductance (junction capacitance). 
We define the charging (inductive) energy of the Josephson junction capacitor (superinductor) as $E_{CJk}=e^2/2C_{Jk}$ ($E_{Lk}=(\frac{\Phi_0}{2\pi})^2/L_{k}$).
These expressions let us further write 
\begin{align}
    Y &= \frac{ 2 Y_{1} Y_{2}}{Y_{1} + Y_{2}},
\end{align}
with $Y \in \{E_{L}, E_{CJ} \}$. Finally, the Josephson energy satisfies 
\begin{align}
    E_{J} &=   \frac{E_{J1} - E_{J2}}{ E_{J1} + E_{J2}}. 
\end{align}
Similarly to the $0-\pi$ qubit, $\cos2\phi$ circuit consists of three degrees of freedom: $\theta$, $\phi$ and $\zeta$, with their respective conjugates defined as $\noper_{\theta}$, $\noper_{\phi}$ and $\noper_{\zeta}$. 
The quantity $n_{g}$ represents the charge offset of the $\noper_{\theta}$ variable. 
We stress that the labels and notation utilized by \scqubits for the degrees of freedom and some of the circuit parameters differs slightly from \cite{Smith2020}. The following table outlines how to convert between the two:
\vspace*{3mm}
{
	\noindent
	\small
	\rowcolors{1}{gray!10}{white}
	\arrayrulecolor{deepblue} 
	\setlength{\tabcolsep}{0pt}
	\setlength\extrarowheight{5pt}
	\begin{tabular}{p{0.35\columnwidth} >{\raggedright\arraybackslash}p{0.65\columnwidth}}
		\rowcolor{lightblue}
        \textbf{\scqubits}                       & \textbf{Reference \cite{Smith2020}}                                                   \\\hline
		$\zeta$ &  $\theta$ \\
		$\theta$ &  $\varphi$ \\
        $\phi$ &  $\frac{\phi}{2}$ \\
        $E_{C}$ &  $x \epsilon_{C}$ \\
        $E_{CJ}$ & $ \epsilon_{C}$  \\
        $E_{J}$ & $ \epsilon_{J}$  \\
        $E_{L}$ & $ \epsilon_{L}$  
	\end{tabular}
}
\vspace*{3mm}

%In particular we have $(E_\text{L2}-E_\text{L1})/(E_\text{L2}+E_\text{L1}), E_\text{L} = 2E_\text{L1}E_\text{L2}/(E_\text{L1}+E_\text{L2})$
%for inductive energies,
%:math:`dC_\text{J} =
%(E_\text{CJ2}-E_\text{CJ1})/(E_\text{CJ1}+E_\text{CJ2}), E_\text{CJ} =
%2E_\text{CJ1}E_\text{CJ2}/(E_\text{CJ1}+E_\text{CJ2})` for charging energies, and
%:math:`dE_\text{J} =
%(E_\text{J1}-E_\text{J2})/(E_\text{J1}+E_\text{J2}), E_\text{J} =
%(E_\text{J1}+E_\text{J2})/2` for junction energies.

To numerically diagonalize the Hamiltonian of the $\cos2\phi$ qubit, the harmonic basis are used for both the $\phiop$ and $\zetaop$ variables (see the discussion of the Fluxonium qubit in \ref{app:fluxonium} for more details), while the charge basis are used for the $\thetaop$ variable (see \ref{app:transmon}). When instantiating an \scqubits object corresponding to this qubit, the user needs to specify cutoffs for basis states described above (this is done using \pyl{phi_cut}, \pyl{zeta_cut}, and \pyl{ncut}), which need to be chosen large enough so that convergence is achieved.

An instance of the $\cos2\phi$ qubit can be initialized as follows
\begin{lstlisting}
cos2phi_qubit = scq.Cos2PhiQubit(
    EJ = 15.0,
    ECJ = 2.0,
    EL = 1.0,
    EC = 0.04,
    dCJ = 0.0,
    dL = 0.6,
    dEJ = 0.0,
    flux = 0.5,
    ng = 0.0,
    ncut = 7,
    phi_cut = 7,
    zeta_cut = 30
)
\end{lstlisting}

\end{document}